\documentclass{jfm}

\usepackage{graphicx}
\usepackage{natbib}
\usepackage{amsmath,epsfig}
\usepackage{url}

\ifCUPmtlplainloaded \else
  \checkfont{eurm10}
  \iffontfound
    \IfFileExists{upmath.sty}
      {\typeout{^^JFound AMS Euler Roman fonts on the system,
                   using the 'upmath' package.^^J}%
       \usepackage{upmath}}
      {\typeout{^^JFound AMS Euler Roman fonts on the system, but you
                   dont seem to have the}%
       \typeout{'upmath' package installed. JFM.cls can take advantage
                 of these fonts,^^Jif you use 'upmath' package.^^J}%
       \providecommand\upi{\pi}%
      }
  \else
    \providecommand\upi{\pi}%
  \fi
\fi

\ifCUPmtlplainloaded \else
  \checkfont{msam10}
  \iffontfound
    \IfFileExists{amssymb.sty}
      {\typeout{^^JFound AMS Symbol fonts on the system, using the
                'amssymb' package.^^J}%
       \usepackage{amssymb}%
       \let\le=\leqslant  \let\leq=\leqslant
         
      }{}
  \fi
\fi

\ifCUPmtlplainloaded \else
  \IfFileExists{amsbsy.sty}
    {\typeout{^^JFound the 'amsbsy' package on the system, using it.^^J}%
     \usepackage{amsbsy}}
    {\providecommand\boldsymbol[1]{\mbox{\boldmath $##1$}}}
\fi

\newsavebox{\astrutbox}
\sbox{\astrutbox}{\rule[-5pt]{0pt}{20pt}}

\def\build#1_#2^#3{\mathrel{\mathop{\kern 0pt#1}\limits_{#2}^{#3}}}
\def \E{ \mathbb E  }
\def \R{ \mathbb R  }

\newcommand\diff{\mathrm{d}}
\newcommand\e{\mathrm{e}}

\newcommand\tr{\mbox{tr}}

\newcommand\bs[1]{\boldsymbol{#1}}
\newcommand\bu{\boldsymbol{u}}
\newcommand\bx{\boldsymbol{x}}
\newcommand\by{\boldsymbol{y}}
\newcommand\bz{\boldsymbol{z}}
\newcommand\bW{\boldsymbol{W}}
\newcommand\bell{\boldsymbol{\ell}}
\newcommand\bh{\boldsymbol{h}}

\newcommand\tX{\mathsfbi{X}}
\newcommand\tA{\mathsfbi{A}}
\newcommand\tS{\mathsfbi{S}}

\title[A dissipative random velocity field for fully developed fluid turbulence]{A dissipative random velocity field for fully developed fluid turbulence}

\author[ ]%
{Rodrigo M. Pereira$^{1,2}$, Christophe Garban$^{3}$, Laurent Chevillard$^1$}

\affiliation{$^1$Univ Lyon, Ens de Lyon, Univ Claude Bernard, CNRS, Laboratoire de Physique \\
$^2$CAPES Foundation, Ministry of Education of Brazil, Bras\'ilia/DF 70040-020, Brazil\\
$^3$Universit\'e de Lyon, Institut Camille Jordan, 43 blvd. du 11 novembre 1918, F-69622 Villeurbanne cedex, France}
\date{}

\begin{document}

\maketitle

\begin{abstract}
We investigate the statistical properties, based on numerical simulations and analytical calculations, of a recently proposed stochastic model for the velocity field of an incompressible, homogeneous, isotropic and fully developed turbulent flow. A key step in the construction of this model is the introduction of some aspects of the vorticity stretching mechanism that governs the dynamics of fluid particles along their trajectories. An additional further phenomenological step aimed at including the long range correlated nature of turbulence makes this model dependent on a single free parameter $\gamma$ that can be estimated from experimental measurements. We confirm the realism of the model regarding the geometry of the velocity gradient tensor, the power-law behaviour of the moments of velocity increments (i.e. the structure functions) including the intermittent corrections and the existence of energy transfer across scales. We quantify the dependence of these basic properties of turbulent flows on the free parameter $\gamma$ and derive analytically the spectrum of exponents of the structure functions in a simplified non dissipative case. A perturbative expansion in power of $\gamma$ shows that energy transfer, at leading order, indeed take place, justifying the dissipative nature of this random field.

\end{abstract}

\normalsize


\section{Introduction}

Fluid turbulence is an archetypal phenomenon belonging to out-of-equilibrium and non-linear classical physics. Starting probably with the work of Reynolds, the complex and multiscale nature of turbulent velocity fluctuations is usually apprehended in a statistical way. In this spirit, \cite{Kol41} proposed in his seminal article a dimensional based argument explaining the spatial two-point correlation structure of velocity fluctuations, i.e. the 2/3-law \citep[see for instance classical textbooks][]{Bat53,TenLum72,Fri95,Pop00}, as it was observed in early experimental measurements of laboratory flows. Furthermore, he derived rigorously from the Navier-Stokes equations, using the stationary solution of the von K\'arm\'an and Howarth equation, the behaviour at infinite Reynolds number and vanishing scale of the third order moment of velocity increments, known as the 4/5-law \citep{Fri95}, reminiscent of a non vanishing mean energy transfer across scales. This gives solid ground to the following phenomenology of three dimensional homogeneous, isotropic and incompressible turbulence: energy is injected at large scale $L$, the so-called integral length scale (typically the mesh size of a grid generated turbulence in a wind tunnel, or the typical scale of propellers, etc.), that is then transferred to smaller scales via a direct cascading process, until it is dissipated by viscosity. 

In this context, a more complete picture could be given while proposing a \textit{stochastic} representation  of a velocity field able to reproduce in probability law the formerly described spatial structure of turbulence. In other words, we ask whether it is possible to build a random vector field $\bu$, incompressible, statistically homogeneous and isotropic, seen as a statistically stationary solution of the Navier-Stokes equations, that reproduces in particular the observed  2/3 and 4/5-law. 

The very first idea would be to consider a Gaussian approximation. This was first considered by Kolmogorov himself, and the respective process belongs to the more general class of fractional Brownian motions \citep{ManVan68}. Unfortunately, such a Gaussian model fails to reproduce the observed and derived mean energy transfer encoded in the third-order moment of the velocity increments, as previously mentioned. However, an underlying Gaussian velocity field is an appealing starting point, and we will see in the following how to modify it in order to obtain a more realistic picture that includes energy transfer.

Hereafter, we consider homogeneous, isotropic and incompressible velocity fields in three dimensional space, i.e. $\bu^\epsilon(\bx) = (u^\epsilon_i(\bx))_{1\le i \le 3}$ and $\bx\in \mathbb R^3 $, with $\epsilon>0 $ a regularizing scale that plays the role, in a schematic way, of the Kolmogorov dissipative length scale. In a Gaussian framework, let us call $\bu^{g,\epsilon}$ such a field. To fully determine this Gaussian velocity field, we have to prescribe its velocity components covariance, that will in particular take into account the self-similar law of Kolmogorov (i.e. the $\frac{2}{3}$-law). Homogeneity, isotropy, incompressibility and the self-similar property can be modelled as a stochastic integral \citep{RobVar08} in the following way
\begin{equation}\label{eq:GaussianField}
\bu^{g,\epsilon}(\bx) = -\int_{\mathbb R^3} \varphi_L(\bx-\bz)\frac{\bx-\bz}{|\bx-\bz|^{5/2-H}_\epsilon}\wedge  \,\bW(\diff z) \mbox{ ,}
\end{equation}
where $\bW(\diff z)=(W_1(\diff z),W_2(\diff z),W_3(\diff z))$ is a zero-average vector Gaussian white noise whose components are independent with variance equal to the infinitesimal volume $\diff z$, and $\varphi_L(\bx)$ is a large-scale cut-off of characteristic extension $L$ (i.e. the integral length scale) ensuring a finite variance of this random velocity field. It is chosen radially symmetric to ensure isotropy, that is, for any vector $\bx$,  $\varphi_L(\bx)=\varphi_L(|\bx|)$. The singular kernel $\bx/|\bx|_\epsilon$ is regularized over the scale $\epsilon$ such that $|\bx|_\epsilon$ is proportional to $\epsilon$ when $|\bx|\to 0$ (see section \ref{sec:NumProc} and \citealt{RobVar08} for further details). This gives differentiability to the vector field $\bu^{g,\epsilon}(\bx)$ for any $\epsilon>0$. The vector product $\wedge$ entering equation (\ref{eq:GaussianField}) recalls the structure of the Biot-Savart law and ensures incompressibility (i.e. a divergence-free vector field). Using standard rescaling techniques (as done in \citealt{RobVar08}, using similar techniques from \citealt{ManVan68}), it can be shown that the limiting process $\bu^{g}=\lim_{\epsilon\to 0}\bu^{g,\epsilon}$, which corresponds in a turbulent context to the limit of vanishing viscosity, is a finite variance random vector field when the so-called Hurst (or H\"older) exponent $H$ is strictly positive, i.e. $H>0$. Let us introduce the \textit{velocity increments} in order to make a connection with turbulence phenomenology and comment on the free parameter $H$ entering the definition of the field from equation (\ref{eq:GaussianField}). As is usually done in the turbulence literature \citep[see for instance classical textbooks such as][]{Bat53,Fri95}, it is convenient to define the longitudinal $\delta_\ell^\parallel u^\epsilon$ and transverse $\delta_\ell^{\perp} u^\epsilon$ velocity increments. Note $u^\epsilon_{\parallel}$ and $u^\epsilon_{\perp}$, the projections of the vector $\bu^\epsilon$ onto the direction of $\ell$ and, correspondingly, onto any perpendicular direction. The longitudinal and transverse velocity increments are given by 
\begin{equation}\label{eq:DefVIs}
\delta_\ell^{\parallel} u^\epsilon(\bx) = u^\epsilon_{\parallel}(\bx+\bell/2)-u^\epsilon_{\parallel}(\bx-\bell/2) \,\mbox{ and } \,\delta_\ell^{\perp} u^\epsilon (\bx) = u^\epsilon_{\perp}(\bx+\bell/2)-u^\epsilon_{\perp}(\bx-\bell/2)\mbox{ .}
\end{equation}
It can be shown, for $0<H<1$, that the limiting Gaussian random field $\bu^g$, as we will recall in this article, is scale-invariant in the sense that moments of velocity increments, the so-called structure functions, behave as power-laws, i.e. for $q\in \mathbb N$,
$$ \mathbb E \left[(\delta_\ell^{\parallel} u^g)^{2q}\right] \build{\sim}_{\ell\rightarrow 0}^{} C_{2q}^{g,\parallel}\ell^{2qH} \, \mbox{ and } \, \mathbb E  \left[(\delta_\ell^{\perp} u^g)^{2q}\right] \build{\sim}_{\ell\rightarrow 0}^{} C^{g,\perp}_{2q}\ell^{2qH},$$
with $C_{2q}^{g,\parallel}$ and $C_{2q}^{g,\perp}$ two strictly positive constants that are universal in the sense that they depend on the order $q$ and only on the value of the cut-off function at the origin $\varphi_L(0)$ (and not on its entire shape). Based on dimensional arguments, Kolmogorov phenomenology predicts $H=1/3$, that is $\mathbb E (\delta_\ell^{\parallel} u^g)^{2} \build{\sim}_{\ell\rightarrow 0}^{} C_{2}^{g,\parallel}\ell^{\frac{2}{3}}$, justifying the terminology of the 2/3-law. We gather the proofs in appendix \ref{ann:GF}. Roughly speaking, the scale-invariance property comes from the singular power-law shape of the kernel entering the definition of the Gaussian velocity field in equation \ref{eq:GaussianField}. We can see that, as far as Kolmogorov's 2/3-law is concerned, we can give a clear meaning (equation \ref{eq:GaussianField}) to a stochastic representation of a turbulent homogeneous, isotropic and incompressible velocity field. Unfortunately, such a representation is too naive to reproduce the 4/5-law, that requires a non vanishing (strictly negative) third order moment of velocity increments, whereas the Gaussian random field (\ref{eq:GaussianField}) is such that 
$$ \mathbb E  \left[(\delta_\ell^{\parallel} u^g)^{3}\right] =\mathbb E  \left[(\delta_\ell^{\perp} u^g)^{3}\right] =0.$$

As noted, the Gaussian velocity field (\ref{eq:GaussianField}) fails to give a realistic picture of K41 phenomenology since the Gaussian structure leads to vanishing odd-order correlators and thus, to a vanishing mean energy transfer across scales. Furthermore, higher-order even correlators are themselves poorly predicted. This intrinsic non-Gaussianity of the small scales was observed in early experimental measurements \citep{Bat53,MonYag71}. This was taken into account by \cite{Kol62} and \cite{Obu62} (hereafter referred to as KO62) while refining the K41 theory and setting a very peculiar statistical structure of the dissipation field. This is the so-called intermittency, i.e. multifractal phenomenon (see \citealt{Fri95} for a review on the subject).

Various approaches were developed in the past to provide a stochastic representation of such a dissipation field, starting probably with the \emph{discrete} cascades models initiated by the Russian school (\citealt{MonYag71}, see also \citealt{BenBif93,ArnBac98}). We will prefer here to follow a continuous version of these discrete cascade models that ensures homogeneity, known as the limit-lognormal model of \cite{Man72}, rigorously studied in the framework of \emph{Gaussian multiplicative chaos} by \cite{Kah85} (see \citealt{RhoVar14} for recent developments on this matter). The aim is to model the dissipation field as a lognormal process with a long range correlation structure of the fluctuations, as observed in experiments \citep{MonYag71,GagHop79,AntPha81}. Gaussian multiplicative chaos consists in defining such a scalar lognormal process as the exponential of a Gaussian field $X(\bx)$ with logarithmic covariance, i.e. $\E [X(\bx)X(\by)]\sim \ln (L/|\bx-\by|)$, with $L$ being the integral length scale. It is then possible to give a clear meaning to the scalar field $e^{\mu X}$ \citep{Kah85}, where $\mu$ is a dimensionless free parameter of the theory. In particular, in three dimensional space, it is then possible to show that the local average of this scalar field $\varepsilon_\ell \propto (1/\ell^3)  \int_{|\bx-\by|<\ell} e^{\mu X(\by)}dy$ over a ball of size $\ell$ is a well-posed random field whose moments are scale invariant in the sense that $\E\varepsilon_\ell^q$ behaves as $\ell^{\tau_q}$, with $\tau_q = (\mu^2/2)q(1-q)$ a non-linear (quadratic) function of the order $q$. As we said, such a construction is defined up to a dimensionless free parameter $\mu$ known as the intermittency coefficient that can be precisely estimated on experimental signals (see for instance \citealt{CheCas12} and references therein).

Until now, the statistical properties of turbulence that we have mentioned concern mainly the fluctuations of the longitudinal velocity profile which is accessible with traditional experimental techniques, hot-wire anemometry in particular, and do not characterize the vector nature of the velocity field. For example, at this stage, nothing is said about the peculiar correlation structure of the components of the velocity gradients tensor $A_{ij}=\partial_ju_i$. We are thus asking if, furthermore, it is possible to build up a differentiable velocity field (at a finite $\epsilon$) consistent with two important properties of the velocity gradient tensor that are (i) the teardrop shape of the joint density of the invariants $Q=-\frac{1}{2}\,\tr(\tA^2)$ and $R=-\frac{1}{3}\,\tr(\tA^3)$ \citep{Tsi01,Wal09,Men11} and (ii) the preferential alignment of the vorticity vector $\bs{\omega}=\nabla \wedge \bu$  with the eigenframe of the rate-of-strain matrix.

Going beyond the Gaussian approximation (\ref{eq:GaussianField}) is a difficult matter since the mathematical theory of non-Gaussian processes is far more sophisticated. In this direction, some recent attempts by \cite{Cag07} and \cite{HedSch14} are interesting but it is not clear whether these vector fields exhibit energy transfer. Let us also mention the iterative procedure of \cite{RosMen08} that gives a realistic picture but which is not explicit, making analytical results out of reach at the present time. In a one-dimensional context, several models have been proposed in the literature in order to apply the discrete cascade models to reproduce synthetically the observed fluctuations of longitudinal velocity profiles, including a model for energy transfer \citep{JunLat94} with additional parameters and propositions to extend to spatio-temporal \citep{BifBof98} and L\'evy-based \citep{SchCle04} stochastic representations.
In a different spirit, \cite{NawPei06} propose to reconstruct velocity time series, starting from a time series at a given (small) scale and assuming a Markov property in scale. As far as we know, \cite{RobVar08} are the first to have proposed a compressible velocity field with non-symmetrical probability laws. To generalize their approach to incompressible velocity fields, they propose to modify the Gaussian field (\ref{eq:GaussianField}) in order to include energy transfer and intermittency effects. This was done while disturbing the vector white noise field $\bW$ by the scalar multifractal measure given by the multiplicative chaos. Unfortunately, for symmetry reasons, \cite{RobVar08} show that this incompressible intermittent velocity field has a vanishing mean energy transfer. It is tempting to think that the present picture is too heuristic to represent the complex local structure of turbulence. A further step in this direction was proposed by \cite{CheRob10} in which the Euler equations, and more precisely the vorticity stretching mechanism, is used in order to motivate the exponentiation of a homogeneous field of isotropic symmetric trace-free Gaussian matrices, that eventually lead to energy transfer. 

Let us recall how to include some aspects of the vorticity stretching mechanism in the present picture. The Euler equations reads, in the vorticity formulation,
$$ \frac{D\bs{\omega}}{Dt}=\frac{\partial \bs{\omega}}{\partial t} + (\bu\cdot\nabla)\bs{\omega} = \tS\bs{\omega}\mbox{ ,}$$
where $\bu$ is the velocity field, solution of the Euler equation and given by the Biot-Savart law, i.e.
$$ \bu(\bx) = -\frac{1}{4\upi}\int \frac{\bx-\bz}{|\bx-\bz|^3}\wedge \bs{\omega}(\bz)\diff z,$$
and the deformation field $\tS(\bx)$ is defined as the symmetric part of the velocity gradient tensor, namely $\tS=\frac{1}{2}\left(\tA+\tA^\top\right)$, where $^\top$ stands for matrix transpose. In incompressible flows, the deformation field is fully determined by the vorticity field and the explicit form reads \citep{Con94,MajBer02}
\begin{equation}\label{eq:DeformationPV} 
\tS(\bx)= \frac{3}{8\upi}\mbox{P.V.}\int\left[\frac{(\bx-\bz)\otimes [(\bx-\bz)\wedge \bs{\omega}(\bz)]}{|\bx-\bz|^5} +\frac{ [(\bx-\bz)\wedge \bs{\omega}(\bz)]\otimes(\bx-\bz)}{|\bx-\bz|^5}\right]\diff z\mbox{ ,}
\end{equation}
where the integral is understood as a Cauchy Principal Value (P.V.) and $\otimes$ is the tensor product, i.e. $(\bx\otimes \bz)_{ij} = x_iz_j$. The first underlying idea of \cite{CheRob10} is to study the implication of a linearization of the previous formulation of the Euler equation on the velocity field generated by the stretching of an initial Gaussian vorticity field (with a K41 structure) by the initial deformation field. This first motivates the use of the exponentiation of a Gaussian random field of symmetric matrices, although it was not expected from this short time study of the Euler equations to reproduce the peculiar intermittent nature of the velocity field. This structure was introduced by hand using the Gaussian multiplicative chaos that is naturally obtained while modifying the integration kernel of the deformation field (equation \ref{eq:DeformationPV}). This heuristic procedure, motivated by the short time dynamics of the Euler equations, leads to the following proposition of a velocity field representing a realistic local structure of turbulence:
\begin{equation}\label{eq:MultifractalField}
\bu^\epsilon(\bx) = -\int_{\mathbb R^3} \varphi_L(\bx-\bz)\frac{\bx-\bz}{|\bx-\bz|^{5/2-H}_\epsilon}\wedge  \, \e^{\gamma \tX^\epsilon(\bz)}\bW(\diff z) \mbox{ ,}
\end{equation}
where $\tX^\epsilon(\bz)$ is an isotropic trace-free symmetric random matrix, whose structure recalls the one of the deformation field (\ref{eq:DeformationPV}), given explicitly by a tensor Wiener integral that we will specify later. The non dimensional constant $\gamma$ governs the level of intermittency. Let us finally remark that a crucial step of this construction, as dictated by the short-time dynamics of the Euler equations, is the intrinsic dependence of this statistically isotropic matrix $\tX^\epsilon$ on the vector white noise $\bW$. We can see, given a Hurst exponent that we will take to be $H=1/3$ to be consistent with K41 phenomenology, that the proposed stochastic model (\ref{eq:MultifractalField}) does depend on a single free parameter $\gamma$ that can be determined empirically. Therefore, if this vector field is to provide a realistic picture of the local structure of turbulence, this unique free parameter should govern at the same time both the intermittency phenomenon and the physics of the energy transfer, which is, as far as we know, a new type of relationship between these phenomena. We will indeed derive from a perturbative approach (section \ref{sec:EnerTrans}) that the third order moment of velocity increments is proportional to the scale, with a multiplicative factor that is itself proportional to this free parameter $\gamma$.

The purpose of this article is to go beyond the results obtained by \cite{CheRob10} in which the field (\ref{eq:MultifractalField}) has been proposed for the first time and studied mostly numerically for a single value of the intermittency coefficient $\gamma$ representing in a satisfactory manner the statistical properties of turbulence. As we will see in the following quick description of the various sections of the article, the proposed new material include (i) an extensive numerical study of the statistical properties of the velocity field at the smallest $\epsilon$ resolutions we were able to reach, for several values of the free parameter $\gamma$, (ii) an analytical derivation of the spectrum of exponents of the structure functions in the asymptotic limit of vanishing resolutions $\epsilon\to 0$ of a simplified ersatz of the field named $\bu^{ind}$ and (iii) a perturbative approach for small $\gamma$ able to capture some aspects of the energy transfer taking place while reconsidering the field $\bu^\epsilon(\bx)$ (equation \ref{eq:MultifractalField}).

In section \ref{sec:NotBas}, we set our notation and define the field of random matrices $\tX^{\epsilon}$.

In section \ref{sec:NumProc}, we describe the numerical procedure in order to obtain realizations of the velocity field (\ref{eq:MultifractalField}). In short, $\bu^\epsilon$ is simulated in a periodic box of size $(2\upi)^3$. We rely then on the discrete Fourier transform to perform the convolutions. The matrix exponential is evaluated at each point of space using a Pad\'e approximant with scaling and squaring. The Fast Fourier Transform (FFT) algorithm is used in its fully parallel form. We study then the numerical properties of the velocity field based on realizations up to $2048^3$ collocation points.

In section \ref{sec:NumA}, we use these numerical simulations to compute the joint density of the invariants $Q$ and $R$ at various intermittency coefficients $\gamma$ and discuss their comparison with what is obtained in laboratory and numerical flows. Similarly, we show the preferential alignment of vorticity with the intermediate eigendirection of the eigenframe of the deformation, and quantify its dependence on $\gamma$.

Section \ref{sec:Intermittency} is devoted to a joint numerical and analytical study of the intermittency phenomenon observed in the velocity field (\ref{eq:MultifractalField}). We will indeed observe that this field is intermittent (in a sense that we will make precise in the devoted section), and its level of intermittency is given in terms of the coefficient $\gamma$. A rigorous derivation of the behaviour of the structure functions of the velocity field is mathematically very demanding, and even obtaining the variance of the components is a difficult task. The reason is related to the strong correlation between the exponentiated Gaussian field of matrices $\tX^\epsilon$ and the underlying vector white noise $\bW$. To obtain analytical results, we study an \textit{ersatz}, which has the same structure as the proposed field (\ref{eq:MultifractalField}) but assuming the independence of the matrix $\tX^\epsilon$ and vector $\bW$ fields. We will call this case the \textit{independent} case and note the respective velocity field $\bu^{ind,\epsilon}$. We will show in section \ref{sec:Intermittency} that indeed, when properly renormalized, the velocity field $\bu^{ind,\epsilon}$ converges towards a finite-variance process $\bu^{ind}$ when $ \epsilon \to 0$ and we will compute its respective structure functions, obtaining
$$ \mathbb E (\delta_\ell^{\parallel} u^{ind})^{2q} \build{\sim}_{\ell\rightarrow 0}^{} C_{2q}^{ind,\parallel}\frac{\ell^{2qH-2q(q-1)\gamma^2}}{\left(\ln \frac{1}{\ell}\right)^{q-1}} \, \mbox{ and } \, \mathbb E (\delta_\ell^{\perp} u^{ind})^{2q} \build{\sim}_{\ell\rightarrow 0}^{} C^{ind,\perp}_{2q}\frac{\ell^{2qH-2q(q-1)\gamma^2}}{\left(\ln \frac{1}{\ell}\right)^{q-1}},$$
with $C_{2q}^{ind,\parallel}$ and $C_{2q}^{ind, \perp}$ two strictly positive constants. Note that, asymptotically, higher order longitudinal and transverse structure functions share similar scaling behaviour. Note also that we do not obtain perfect power-laws since an additional logarithmic factor appears in the asymptotic behaviour. This factor is related to the matrix nature of the chaos and was already observed in \cite{CheRho13}. This former scale dependence of structure functions is based on an exact calculation for  $q\le 2$, and has been extended to higher orders $q>2$ based on a conjecture proposed in \cite{CheRho13}. Thus, this field $u^{ind}$ allows us to understand the intermittent corrections to the scaling behaviour with respect to the Gaussian case $\bu^g$. The independence assumption leads on the other side to vanishing third- and more generally odd-orders structure functions, namely
$$ \mathbb E (\delta_\ell^{\parallel} u^{ind})^{3}=0,$$
missing all the physics of energy transfer and showing that the intrinsic correlation between the matrix $\tX^\epsilon$ and vector $\bW$ fields in the velocity field $\bu^{\epsilon}$ (equation \ref{eq:MultifractalField}) is crucial to reproduce non vanishing third order moment. Nonetheless, we show numerically that, to fourth order, the ersatz $\bu^{ind,\epsilon}$ and the full velocity field (\ref{eq:MultifractalField}) share similar intermittent properties. Furthermore, an analytical study that takes into account finite-scale corrections is performed in order to interpret with high precision the numerical results. From the behaviour of the velocity increment flatness that we will define later on (equation \ref{eq:Flatnesses}), taking into account non trivial finite-scale corrections, we are led to propose the very particular value $\gamma^2=0.067$ for turbulent applications in order to be consistent with experimental measurements and numerical simulations.

Section \ref{sec:EnerTrans} is devoted to the physics of energy transfer. As we said, a rigorous study of the statistical properties of the proposed velocity field (\ref{eq:MultifractalField}) is a difficult task. In order to discuss the important physics of the energy transfer, as required by the $4/5$-law \citep{Fri95}, we will rely on a perturbative analysis of this field, at a finite resolution $\epsilon>0$, using the intermittency parameter $\gamma$ as the small parameter, which is indeed the case as far as turbulence is concerned. We show that such a perturbative expansion of the longitudinal third order velocity structure function is given by
$$\E(\delta_\ell^{\parallel} u^\epsilon)^3 =\gamma  \mathcal D_\epsilon(\ell) \ell^{3H} + o_\epsilon(\gamma),$$
where $ o_\epsilon(\gamma)$ stands for a term that depends on $\epsilon$ but depends on a higher power of $\gamma$ than 1 (typically this term is of order $\gamma^3$ by symmetry). We are then able to show that the dominating term linked to $\mathcal D_\epsilon(\ell)$ converges when $\epsilon\to 0$ towards a non trivial function $\mathcal D(\ell)$ which is such that
$$ \lim_{\epsilon\to 0}  \mathcal D_\epsilon(\ell)= \mathcal D(\ell)\build{\longrightarrow}_{\ell \to 0}^{}\mathcal D(0) = D,$$
with $D$ a constant. Numerical simulations show indeed such a linear behaviour, with $D<0$, of the third order structure function with both the intermittency coefficient $\gamma$ and the scale $\ell$ when $\gamma$ is small and when we use the Hurst of K41, namely $H=1/3$. This shows, up to first order in $\gamma$, that the proposed velocity field (\ref{eq:MultifractalField}) exhibits energy transfer according to Kolmogorov phenomenology.

We gather in section \ref{sec:Conclusion} our conclusion and perspectives.





\normalsize

\section{Notations and basic properties of the velocity field}\label{sec:NotBas}

In what follows, $\delta_{ij}$ will denote the Kronecker delta and $\epsilon_{ijk}$ the Levi-Civita symbol. We adopt Einstein's convention  of sum over repeated indices, unless explicitly stated, and we note that $\epsilon_{ijk}\epsilon_{ipq}=\delta_{jp}\delta_{kq}-\delta_{jq}\delta_{kp}$. 

The full vector field (\ref{eq:MultifractalField}) reads, with index notation,
 \begin{equation}\label{eq:MultifractalFieldIndices}
u_i^\epsilon(\bx) =\int\phi_{ik}^\epsilon(\bx-\bz)\left(\e^{\gamma \tX^\epsilon(\bz)}\right)_{kl}W_l(\diff z),
\end{equation}
where the kernel $\phi_{ik}$ encodes the structure of the underlying Gaussian velocity field (\ref{eq:GaussianField}) and is given by
$$ \phi_{ik}^\epsilon(\bx) =-\epsilon_{ijk}\varphi_L(\bx)\frac{x_j}{|\bx|_\epsilon^{\frac{5}{2}-H}}, $$ 
and the following matrix field built from the very same vector white noise $\bW$ that enters the construction of the underlying structure:
\begin{equation}\label{eq:Xepsilon}
\tX^\epsilon(\bx) = \sqrt{\frac{15}{32\upi}}\int_{|\bx-\by|\le L} \frac{\bx-\by}{|\bx-\by|^{7/2}_\epsilon}\otimes [(\bx-\by)\wedge \bW(\diff y)]+ [(\bx-\by)\wedge \bW(\diff y)] \otimes \frac{\bx-\by}{|\bx-\by|^{7/2}_\epsilon},
\end{equation}
which is inspired by the tensor structure of the rate-of-strain matrix $\tS$ (equation \ref{eq:DeformationPV}) that stretches the vorticity vector along its path. We will motivate the use of the multiplicative factor $\sqrt{\frac{15}{32\upi}}$ when we give the variance and covariance of the elements of the matrix $\tX^\epsilon$. At this stage, remark that this matrix is Gaussian since it is defined through a linear operation on a Gaussian measure $\bW$ (equation \ref{eq:Xepsilon}). It is indeed symmetric, and it is easy to check that it is trace free, according to 
$$ \mbox{tr}(\tX^\epsilon)=\sqrt{\frac{15}{32\upi}}\int_{|\bx-\by|\le L} 2\frac{\bx-\by}{|\bx-\by|^{7/2}_\epsilon}\cdot[(\bx-\by)\wedge \bW(\diff y)]=0.$$
The free parameter $\gamma$ entering the velocity field $\bu^\epsilon$ (equation \ref{eq:MultifractalFieldIndices}) plays the same role as the free parameter $\lambda$ used in the field given by equation 12 of \cite{CheRob10}, and their relation is $\gamma^2=(8/3)\lambda^2$.

The Gaussian white vector field $W_i(\bx)$, for $1\le i\le 3$, follows the following rules of calculation. For any suitable deterministic function $f(\bx,\by)$, $(\bx,\by)\in(\mathbb R^3)^2$ such that it is integrable along its diagonal, we have
$$ \E \int f(\bx,\by)W_i(\diff x)= \int f(\bx,\by)\E\left[W_i(\diff x)\right]= 0,$$
and
$$ \E \int f(\bx,\by)W_i(\diff x)W_j(\diff y)= \int f(\bx,\by)\E\left[W_i(\diff x)W_j(\diff y)\right]= \delta_{ij} \int f(\bx,\bx)\diff x.$$

\subsection{Covariance structure of the field of isotropic matrices}

\label{sec:CovFieldMatrix}

Let us first show that the matrix field $\tX^\epsilon$ (\ref{eq:Xepsilon}) is indeed homogeneous and isotropic. Homogeneity of the field $\tX^\epsilon(\bx)$  follows from the convolution with the homogeneous white measure $\bW$. Consider now a rotation matrix $\mathsfbi{R}\in O_3(\mathbb R)$ such that $\mathsfbi{R}\mathsfbi{R}^\top=\mathsfbi{I}$, where $\mathsfbi{I}$ the $3\times 3$ identity matrix. Then, it can be shown that for any rotation matrix $\mathsfbi{R}$, we have $\tX^\epsilon(\bx) \stackrel{law}{ =} \mathsfbi{R}\tX^\epsilon(\bx)\mathsfbi{R}^\top$. The equality in law $\stackrel{law}{=}$ stands for equality in probability. This shows that, in that sense, the matrix field is statistically isotropic.

As an important further characterization of the homogeneous field of matrices $\tX^\epsilon$, we want to obtain its covariance structure, component by component. Let us first remark that all of the elements  $(X^\epsilon_{ij})_{1\le i\le j\le 3}$  are of zero mean, which follows from the definition of the field as a convolution with a zero-mean white noise $ \mathbb E[W_i]=0$. We gather all the proofs of the following results in annex \ref{ann:CorrX}. 

We have seen that the field of matrices $\tX^\epsilon$ is statistically isotropic. Recall that each element is a Gaussian random variable, and $\tX^\epsilon$ is a symmetric matrix. Thus, the covariance structure of its elements is given by the general framework developed in \cite{CheRho13}. We recall several key properties of this random matrix.

The first property of the elements of $\tX^\epsilon$ is the divergence of their variance with the regularizing parameter $\epsilon$. Henceforth, we focus only on the element $X^\epsilon_{11}$ of the matrix. See annex \ref{ann:CorrX} for a general discussion on the statistical behaviour of the other elements. Defining the variance of this element as $\sigma_\epsilon^2$, then it is easy to obtain its asymptotic behaviour when $\epsilon\to 0$ as
\begin{equation}  \label{eq:AsympSigmaEps}
\sigma_\epsilon^2=\mathbb E[(X_{11}^\epsilon)^2] \build{\sim}_{\epsilon\rightarrow 0}^{} \ln\frac{L}{\epsilon}\mbox{ .}\end{equation}
Thus, the variance of the elements of $\tX^\epsilon$ diverge logarithmically with $\epsilon$. This situation is classically encountered in the context of multiplicative chaos (see a review on this topic by \citealt{RhoVar14}). Similarly, the covariance of the element $X_{11}^\epsilon$ can be computed and we find, taking first the limit $\epsilon \to 0$ and then looking for an equivalent at small distances,
\begin{equation}\label{eq:AsympSigmaCorr}
\sigma_{|\bx-\by|}^2 = \lim_{\epsilon\to 0}\E[X^\epsilon_{11}(\bx)X^\epsilon_{11}(\by)] \build{\sim}_{|\bx-\by|\rightarrow 0}^{}\ln\frac{L}{|\bx-\by|} \mbox{ .}
\end{equation}
In other words, the Gaussian random variable $X^\epsilon_{11}$ converges when $\epsilon\to 0$ towards a random Gaussian distribution whose covariance behaves logarithmically at small distances. We remark that the factor $ \sqrt{15/(32\upi)}$ entering the definition of $\tX^\epsilon$ (\ref{eq:Xepsilon}) ensures a unit-factor in front of the logarithmic behaviours seen in equations (\ref{eq:AsympSigmaEps}) and (\ref{eq:AsympSigmaCorr}). The very purpose of the theory of multiplicative chaos \citep{RhoVar14} is to give a meaning to the exponential of such a field.

\subsection{Homogeneity and isotropy}\label{sec:HomIso}

Let us now show that the velocity field (\ref{eq:MultifractalFieldIndices}), defined with the field of matrices $\tX_\epsilon$ (\ref{eq:Xepsilon}), is indeed homogeneous and isotropic. Again, homogeneity of the vector field $\bu^\epsilon$ follows from the convolution with the homogeneous field $\e^{\gamma \tX^\epsilon(\bz)} \,\bW(\diff z)$. Consider again a rotation matrix $\mathsfbi{R}\in O_3(\mathbb R)$. Then, for any rotation matrix $\mathsfbi{R}$, we have $\bu^\epsilon(\bx) \stackrel{\mathrm{law}}{ =} \mathsfbi{R}\bu^\epsilon(\bx)$. Thus, the velocity field is statistically isotropic. As a consequence, the velocity field is of zero-mean, i.e., for any $\epsilon>0$
$$\E \bu^\epsilon = \bs{0}.$$


\section{Numerical procedure}\label{sec:NumProc}

As we will see in the following, a rigorous derivation of the statistical properties of the velocity field $\bu^\epsilon$ is a difficult matter. When analytical results are not possible, we will rely on numerical simulations. To do so, we perform a numerical approximation of $\bu^\epsilon$ in the periodic domain $[0,2\upi]^3$. Define $N$ as the number of collocation points in one direction. We will typically present results for $N=2048$, using a fully parallelized algorithm of the Fast Fourier Transform \citep{FFTW}. The elementary volume is given by $\diff x=(2\upi/N)^3$. Standard algorithms allow to generate $3N^3$ independent realizations of a zero-mean Gaussian variable of variance $\diff x$ in order to define the vector white noise $\bW(\diff z)$. The elements of the matrix exponential entering the construction are calculated with the Expokit tool \citep{expokit} using the irreducible rational Pad\'e approximant. The remaining convolutions are performed in the Fourier space.

We choose as an isotropic cut-off function the following $\mathcal C^\infty$, compactly supported function
$$ \varphi_L(\bx) = \e^{-\frac{|\bx|^2}{L^2-|\bx|^2}}1_{|\bx|\le L},$$
and we will consider the particular value $L=\upi/2$ in order to obtain a couple of integral scales in our simulations. The precise shape of this function is not important, besides its characteristic length scale $L$, and only the large scale statistical quantities such as the variance depend on it. We will see that at small scales, explored as an example by velocity increments, only the value at the origin $\varphi_L(0)$ matters.

As a regularization mechanism, we use the following regularized norm
\begin{equation}\label{eq:RegulNorm} 
|\bx|^2_\epsilon =  |\bx|^2 +\epsilon^2.
\end{equation}
This regularization procedure makes the continuous field $\bu^\epsilon$ (\ref{eq:MultifractalFieldIndices}) differentiable, and divergence free in particular. In the discrete approximation, we cannot choose $\epsilon$ arbitrarily small, since it is bounded from below by the finiteness of the smallest accessible scale $(\diff x)^{1/3} = 2\upi/N$. As mentioned, $\epsilon$ plays the role, in a schematic way, of the Kolmogorov scale, and therefore should depend on the Reynolds number. As far as the Gaussian field is concerned (equation \ref{eq:GaussianField}), we can relate $\epsilon$ to the kinematic viscosity $\nu$ and the Hurst exponent $H$ in such a way that the average dissipation per unit of mass remains finite and strictly positive when $\nu \to 0$ \citep{CheHDR15}. In the following, we will work numerically at a finite viscosity, and we will be interested theoretically in the asymptotic limit $\epsilon\to 0$, which corresponds to the infinite Reynolds number limit. We thus have to take $\epsilon$ greater than $(\diff x)^{1/3} = 2\upi/N$. When $\epsilon\gg (\diff x)^{1/3}$, then the numerical field is smooth and gradients are well approximated. In particular, in standard deviation, the divergence of the field $\mbox{div}(\bu^\epsilon)$ is much smaller than the gradient of one of its components. When $\epsilon\approx (\diff x)^{1/3}$, the numerical field is rough, and gradients are poorly approximated. In other words, the divergence of the field can be of the order of the gradient of one of its components (in standard deviation). We are also interested in simulations where the \textit{inertial range} is wide, i.e. we would like to maximize the ratio $L/\epsilon$. In section \ref{sec:NumA}, since we will focus on velocity gradients, we will use $\epsilon=3(\diff x)^{1/3}$. In the following sections, we will use $\epsilon=(\diff x)^{1/3}$. Once again, the precise regularization procedure is not important as long as $|\bx|_\epsilon$ is of order $\epsilon$ at the origin, and equal to $|\bx|$ at a distance $\gg \epsilon$ from the origin. We can rigorously show that this is the case for the matrix multiplicative chaos \citep{CheRho13}. 

\section{Statistical structure of the velocity gradient tensor}\label{sec:NumA}

\begin{figure}
\begin{center}
\epsfig{file=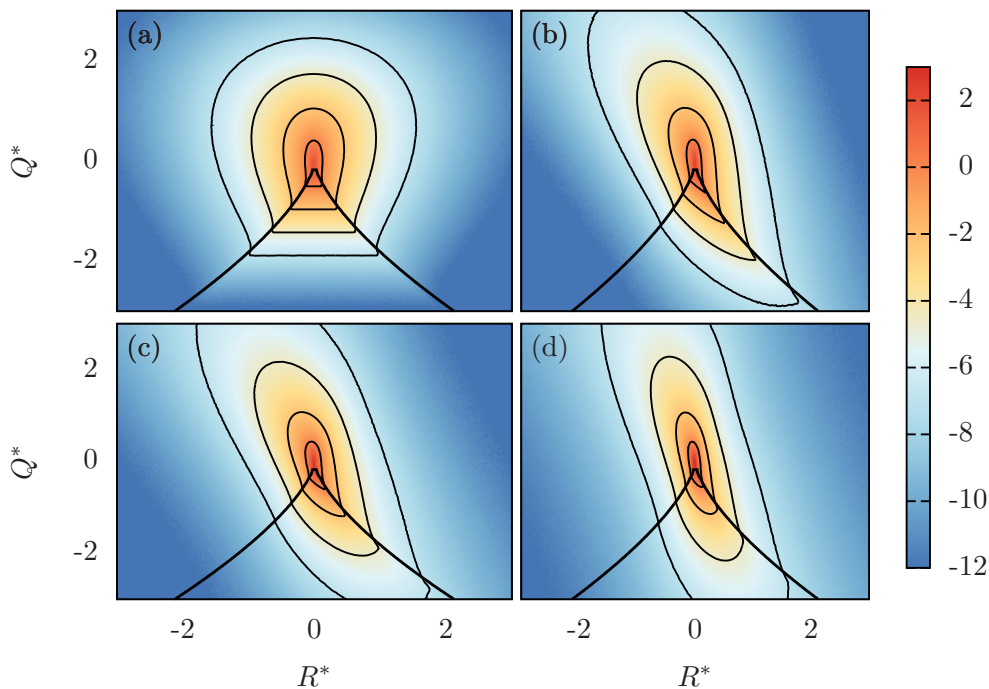,width=13cm}
\end{center}
\caption{\label{fig:RQs} Logarithmic representation of the joint probability density $\mathcal P(Q^*,R^*)$ of $R^* = R/\langle S_{ij}S_{ij}\rangle^{3/2}$ and $Q^*= Q/\langle S_{ij}S_{ij}\rangle$ calculated from the simulation of the field $\bu^{\epsilon}$ using $L=\upi/2$, $N=2048$ and $\epsilon=3(\diff x)^{1/3}$.  (a) $\gamma^2=0$ (Gaussian case), (b) $\gamma^2=0.033$, (c) $\gamma^2=0.067$ and (d) $\gamma^2=0.133$. Contour lines are the same in all cases, logarithmically spaced by a factor of 10, and start at 1 near the origin. The thick line represents the zero-discriminant (or Vieillefosse) line: $(27/4)R^2+Q^3=0$. }
\end{figure}

As an important characterization of the velocity gradient tensor $A_{ij}=\partial_j u_i^\epsilon$, we study its two non vanishing invariants. For instance, the second invariant $Q$ is given by
\begin{equation}\label{eq:DefQ}
Q = -\frac{1}{2}\tr(\tA^2) = \frac{1}{4} |\bs{\omega}|^2-\frac{1}{2} \tr(\tS^2)
\end{equation}
where $\bs{\omega}$ is the vorticity vector and $\tS$ the symmetric part of $\tA$, and can be interpreted as the competition between enstrophy and dissipation (per unit viscosity). Then, positive $Q$ represents rotation-dominated
regions and negative $Q$ dissipation-dominated regions. Analogously, the third invariant $R$ is given by
\begin{equation}\label{eq:DefR}
R= -\frac{1}{3}\tr(\tA^3) = -\frac{1}{4} \omega_iS_{ij}\omega_j-\frac{1}{3} \tr(\tS^3)
\end{equation}
representing competition between \textit{enstrophy production} and \textit{dissipation production}. See \cite{Tsi01,Wal09,Men11} for a discussion on this topic. We simulate the vector field $\bu^\epsilon$ for four different values of intermittency coefficients $\gamma$, with $L=\upi/2$, $N=2048$ and $\epsilon=3(\diff x)^{1/3}$ (see discussion in section \ref{sec:NumProc})  and represent the numerical estimation of the joint density of the invariants $Q$ and $R$ in figure \ref{fig:RQs}.

As is well known, a Gaussian velocity field corresponding to $\bu^{g,\epsilon}$ (\ref{eq:GaussianField}), or equivalently the velocity field $\bu^\epsilon$ (\ref{eq:MultifractalField}) with $\gamma=0$, predicts a joint density of the invariants symmetrical with respect to the $R=0$ line. This is what we obtain in figure \ref{fig:RQs}(a). In figures \ref{fig:RQs}(b,c,d), we study the effect of increasing $\gamma$. We can see that the bigger the value of $\gamma$, the more elongated is the joint density along the right tail of the zero discriminant (or Vieillefosse) line, where $(27/4)R^2+Q^3=0$ \citep{Vie82}. We will see that increasing $\gamma$ corresponds to increasing the level of intermittency. As justified in section \ref{sec:Intermittency}, we choose the very particular value $\gamma^2=0.067$ for turbulence applications whose corresponding joint density of $R$ and $Q$ is displayed in figure \ref{fig:RQs}(c).

\begin{figure}
\begin{center}
\epsfig{file=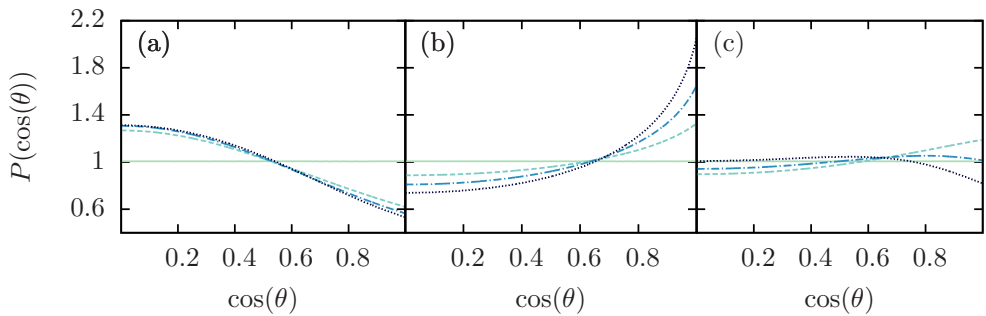,width=13cm}
\end{center}
\caption{\label{fig:Aligns} Probability densities of $\cos(\theta)$, where $\theta$ is the angle between vorticity and the eigenvectors of the rate of strain tensor, estimated from the same numerical simulation as in figure \ref{fig:RQs}. We represent alignments with the eigenvectors associated to the most negative (a), intermediate (b) and most positive (c) eigenvalues, for the Gaussian case $\gamma=0$ (solid lines),  $\gamma^2=0.033$ (dashed), $\gamma^2=0.067$ (dot-dashed) and $\gamma^2=0.133$ (dotted).}
\end{figure}

Another striking property of turbulence is the preferential alignment of vorticity with the strain eigendirection associated to the intermediate eigenvalue. We refer again to \cite{Tsi01,Wal09,Men11} for further discussions. We represent in figure \ref{fig:Aligns} the probability density of the cosine of the angle $\theta$ between vorticity and the eigenvectors of the strain. Figure \ref{fig:Aligns}(b) indicates the preferential alignment of vorticity with the correct eigenvector, as observed already in \cite{CheRob10} for a single value of $\gamma^2=0.067$. Here, we can see that this alignment is governed by the intermittency coefficient $\gamma$: no preferential alignment is observed when $\gamma=0$, as expected from a Gaussian velocity field, and this preferential alignment increases with increasing $\gamma$. We observe also in figure \ref{fig:Aligns}(a) that the density of the preferential orthogonality of vorticity with the eigendirection associated to the smallest (negative) eigenvalue is barely sensitive to $\gamma$, except in the Gaussian case $\gamma=0$. As for the angle between vorticity and the eigendirection associated to the biggest (positive) eigenvalue (figure \ref{fig:Aligns}(c)), as observed in real flows, the density is almost flat, with a slight dependence on the parameter $\gamma$, showing no preferential orientation.

Overall, as far as velocity gradient statistics are concerned, the velocity field $\bu^\epsilon$ predicts non trivial facts of fluid turbulence. In this picture, at least for the range of $\gamma$ values studied, the dependence on this parameter is weak. We will study in the following section the influence of the parameter $\gamma$ on the scaling of structure functions, where it will play a key role.

\section{Numerical and theoretical study of the intermittent properties}\label{sec:Intermittency}

As we have seen in section \ref{sec:NumA}, at a finite $\epsilon$ (or finite Reynolds number), the velocity field $\bu^\epsilon$  (\ref{eq:MultifractalField}) predicts realistic velocity gradient statistics. In particular, at any $\gamma>0$, we reproduce the teardrop shape of the $RQ$-plane (figure \ref{fig:RQs}) and the preferential alignment of vorticity (figure \ref{fig:Aligns}). The precise value of $\gamma$ representing realistic turbulent statistics has not been selected yet. This is the purpose of this section.

\begin{figure}
\begin{center}
\epsfig{file=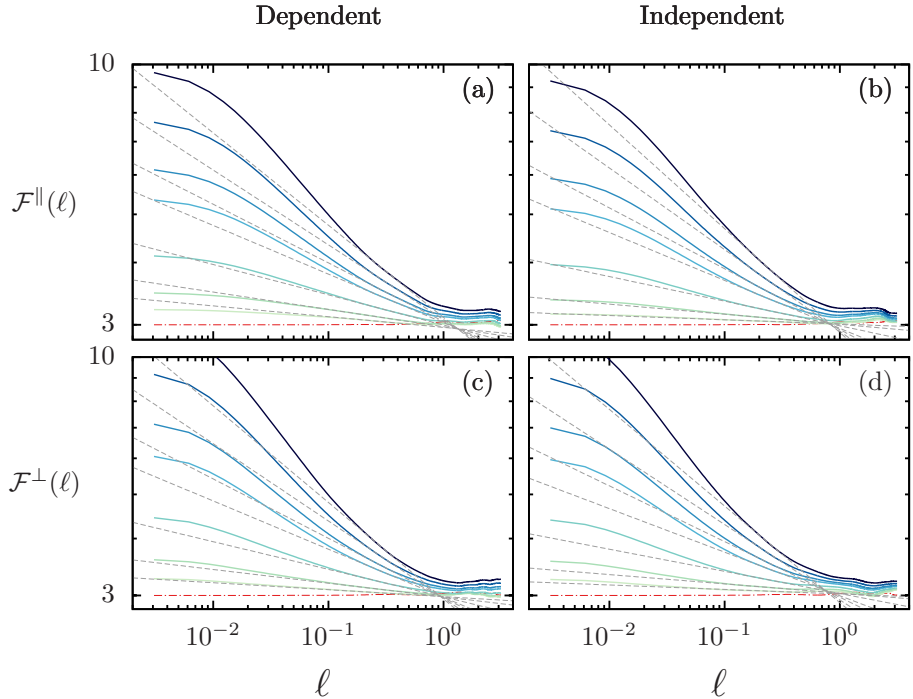,width=12cm}
\end{center}
\caption{\label{fig:flatness}Logarithmic representation of the flatness of longitudinal (top) and transverse (bottom) velocity increments (equation \ref{eq:Flatnesses}) as a function of scales. The parameters of the simulation are $N=2048$, $L=\upi/2$ and $\epsilon=(\diff x)^{1/3}\approx 0.00307$ (see discussion in section \ref{sec:NumProc}). We use several values for the parameter $\gamma^2=0, 0.01, 0.02, 0.04, 0.067, 0.08, 0.10, 0.12$, that give in all cases and any scale $\ell$ an increasing value of the flatness. In (a) (respectively (c)) we represent the flatnesses of the longitudinal (transverse) increments corresponding to the vector field $\bu^\epsilon$ (\ref{eq:MultifractalField}). We represent in a similar way in (b) and (d) the flatnesses but for the velocity field $\bu^{ind,\epsilon}$ (\ref{eq:MultifractalFieldIndicesIndpt}). The power-laws observed at moderate scales are superimposed on the same plots.}
\end{figure}

We focus now on the intermittency phenomenon, and explore the influence of the parameter $\gamma$ on the anomalous scaling of the structure functions \citep{Fri95}. To do so, we will discuss this phenomenon based on the flatness of velocity increments (defined in equation \ref{eq:DefVIs}), namely
\begin{equation}\label{eq:Flatnesses}
\mathcal F^{\parallel,\perp}_\epsilon (\ell) = \frac{\E (\delta^{\parallel,\perp}_\ell u^\epsilon)^4}{[\E (\delta^{\parallel,\perp}_\ell u^\epsilon)^2]^2},
\end{equation}
in both the longitudinal (i.e. $\parallel$) and transverse cases ($\perp$). We perform first simulations of the field for $N=2048$, $L=\upi/2$ and several $\gamma$. We choose $\epsilon=(\diff x)^{1/3}$ in order to maximize the extend of the inertial range, and represent the results of the estimation of the flatness in figure \ref{fig:flatness}.

\subsection{Numerical estimations}\label{sec:IntermittencyNS}

\begin{figure}
\begin{center}
\epsfig{file=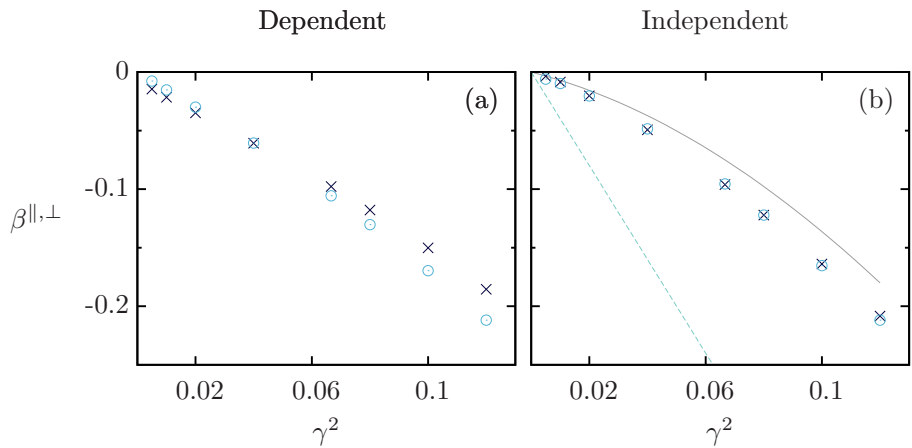,width=12cm}
\end{center}
\caption{\label{fig:flatness_fit} Estimation of the power-law exponents $\beta^{\parallel,\perp}(\gamma)$ defined in \ref{eq:FlatnessesExpo} from the fits performed in figure \ref{fig:flatness} as a function of $\gamma^2$. We represent in figure (a) the results in the dependent case (corresponding to the fits of figures \ref{fig:flatness}(a) and (c)), using the symbols $\times$ for longitudinal, and $\circ$ for the transverse velocity increments. In (b), similar study for the independent case corresponding to the fits of figures \ref{fig:flatness}(b) and (d). We superimpose in (b) the result of our theoretical predictions: (dashed line) the asymptotical prediction $\beta^{\parallel,\perp}(\gamma) = -4\gamma^2$ considering that the exponents $\beta^{\parallel,\perp}(\gamma)$ have been measured at vanishing scales $\ell\to 0$, and (solid line) a prediction that takes into account corrections implied by the finiteness of the scale $\ell$ (see section \ref{sec:FSSFlat}).}
\end{figure}

We display in figure \ref{fig:flatness}(a) (respectively figure \ref{fig:flatness}c) the flatness of the longitudinal (transverse) velocity increments as a function of the scale $\ell$ for selected values of the parameter $\gamma^2$, including the  Gaussian case $\gamma^2=0$. We indeed notice that in the Gaussian case the flatnesses $\mathcal F_\epsilon^{\parallel}$ and $\mathcal F_\epsilon^{\perp}$ do not depend on scale and equal 3. Then, as $\gamma$ increases, for any fixed scale, the flatness increases. We observe a power law with the scale within a limited range, i.e.
\begin{equation}\label{eq:FlatnessesExpo} 
\mathcal F^{\parallel,\perp}_\epsilon (\ell) \build{\sim}_{\epsilon\ll \ell<L}^{} \ell^{\beta^{\parallel,\perp}(\gamma)},
\end{equation}
the regularization over the scale $\epsilon$ polluting a large part of the accessible scales. We represent in figure \ref{fig:flatness_fit}$(a)$ the obtained value for $\beta^{\parallel,\perp}$ as a function of $\gamma^2$. We indeed see that the scaling exponents $\beta^{\parallel,\perp}$ decrease when $\gamma$ increases, showing the augmentation on the level of intermittency \cite{Fri95}. We observe also, in this context, that the level of intermittency of transverse velocity increments is higher than the one observed on longitudinal velocity increments. In laboratory and numerical flows \citep[see for instance][]{CheCas12}, we find a universal behaviour (independent of the flow geometry and the Reynolds number) for the longitudinal case, with $\beta^{\parallel}=-0.1$. Thus, turbulence statistics seem to be well reproduced for a very particular small value of the parameter $\gamma^2 = 0.067$, which was already found in \cite{CheRob10}.

The present velocity field (\ref{eq:MultifractalField}) is an example of random process that exhibits a higher level of intermittency for the transverse case than for the longitudinal case. This is indeed a surprising effect, also observed in real flows (see discussions in \citealt{DhrTsu97,CheSre97,GraHom12}). In our model, only an analytical study could give a clear answer to this observed discrepancy, in particular in the asymptotic limit $\ell \to 0$. Unfortunately, the underlying mathematical structure of this field is subtle, the strong correlation between the field $\tX^\epsilon$ and the white measure $\bW$ is difficult to handle. In the following, we will study both numerically and theoretically an \textit{ersatz} $\bu^{ind,\epsilon}$ of the velocity field $\bu^{\epsilon}$ (\ref{eq:MultifractalField}) that follows the same rules of construction, except that the field of matrices $\tX^\epsilon$ is built independently of the underlying white measure $\bW$. This ersatz is amenable to exact derivations of its statistical properties.

\subsection{The hypothesis of independence}\label{sec:IndptCase}

Consider now the following velocity field
 \begin{equation}\label{eq:MultifractalFieldIndicesIndpt}
u_i^{ind,\epsilon}(\bx) = \frac{1}{c_\epsilon}\int\phi_{ik}^\epsilon(\bx-\bz)\left(\e^{\gamma \tX^\epsilon(\bz)}\right)_{kl}W_l(\diff z),
\end{equation}
where again
$$ \phi_{ik}^\epsilon(\bx) =-\epsilon_{ijk}\varphi_L(\bx)\frac{x_j}{|\bx|_\epsilon^{\frac{5}{2}-H}}, $$ 
and the following \textit{matrix} field:
\begin{equation}\label{eq:GaussSymmMatIndpt}
\tX^\epsilon(\bx) = \sqrt{\frac{15}{32\upi}}\int_{|\bx-\by|\le L} \frac{\bx-\by}{|\bx-\by|^{7/2}_\epsilon}\otimes [(\bx-\by)\wedge \bW'(\diff y)]+ [(\bx-\by)\wedge \bW'(\diff y)] \otimes \frac{\bx-\by}{|\bx-\by|^{7/2}_\epsilon}
\end{equation}
where now the vector noise $\bW'$ is independent of the vector noise $\bW$ of the underlying Gaussian structure (\ref{eq:MultifractalFieldIndicesIndpt}), namely for any $(\bx,\by)\in (\mathbb R^3)^2$, and any components $k$ and $l$, we have $\E [W_k(\bx)W'_l(\by)]=0$. As we show in appendix \ref{ann:IndeptCase}, the field (\ref{eq:MultifractalFieldIndicesIndpt}) needs to be renormalized in order to converge, when $\epsilon \to 0$, towards a finite-variance process. This deterministic normalization constant $c_\epsilon$ (which diverges when $\epsilon\to 0$) is given by
\begin{equation}\label{eq:RenormIndpt} 
c_\epsilon^2 = \frac{1}{3}\E\left[\tr~ \e^{2\gamma \tX^\epsilon}\right]. 
\end{equation}
This is a standard way to renormalize the multiplicative chaos and this situation is well understood as far as multiplicative chaos is concerned \citep{CheRho13,RhoVar14}. We can see that the difference between the full vector field (\ref{eq:MultifractalField}) and the one just mentioned (\ref{eq:MultifractalFieldIndicesIndpt}) is that the matrix free field $\tX^\epsilon$ and the vector white noise $\bW$ are independent. This is a strong simplification but we can get simple exact results.

\subsubsection{Numerical simulations}

We perform similar simulations of the velocity field $\bu^{ind,\epsilon}$ (\ref{eq:MultifractalFieldIndicesIndpt}) as we did for the (dependent) velocity field  $\bu^{\epsilon}$ (\ref{eq:MultifractalFieldIndices}) and compare the estimation of velocity increment flatnesses. We display our results in figures \ref{fig:flatness}(b,d) and \ref{fig:flatness_fit}(b).

Overall, at any value of the parameter $\gamma$, both velocity fields $\bu^{\epsilon}$ and $\bu^{\epsilon,ind}$ share qualitatively similar levels of intermittency. This gives confidence in explaining the intermittent properties of $\bu^{\epsilon}$ (\ref{eq:MultifractalFieldIndices}) using results from the intermittent nature of the velocity field ersatz $\bu^{ind,\epsilon}$ (\ref{eq:MultifractalFieldIndicesIndpt}), which is amenable to analytical derivation. The main difference, as far as intermittency is concerned, comes from the behaviour of the flatness of transverse velocity increments. We observe indeed that in the independent case, the observed power-laws of the flatnesses $\mathcal F_\epsilon^{\parallel}$ and $\mathcal F_\epsilon^{\perp}$ (equation \ref{eq:Flatnesses}) have the same scaling exponent $\beta^{\parallel}(\gamma) =\beta^{\perp}(\gamma)$ (\ref{eq:FlatnessesExpo}), a property that we show in the following sections. Also, as we will see, the velocity field ersatz $\bu^{ind,\epsilon}$ does not exhibit energy transfer. In other words, $\bu^{ind,\epsilon}$ is a good ersatz to study the intermittent nature of the velocity field  $\bu^{\epsilon}$, but fails at describing the physics of energy transfer. This reveals also the great importance of building up the matrix field $\tX^\epsilon$ from the very same white measure $\bW$ in order to predict energy transfer. We will come back to this point in section \ref{sec:EnerTrans}.

\subsubsection{Mean and covariance}

Since $\bW$ and $\tX^\epsilon$ are independent, we easily get (see annex \ref{ann:IndeptCase}) 
\begin{align*}
\E [u_i^{ind,\epsilon}(\bx)] =0.
\end{align*}
Thus, the vector field (\ref{eq:MultifractalFieldIndicesIndpt}) has zero average, as expected from an isotropic vector field. As for the covariance, we find
\begin{align*}
\E [u^{ind,\epsilon}_i(0)u^{ind,\epsilon}_p(\bh)] = (\phi^{\epsilon}_{ik}\star\phi^{\epsilon}_{pk})(\bh),
\end{align*}
where $\star$ is the correlation product defined in equation (\ref{eq:CovGF}). We can see that the covariance structure of the vector field $\bu^{ind,\epsilon}$ (\ref{eq:MultifractalFieldIndicesIndpt}) is the same as the one obtained from the underlying Gaussian field $\bu^{g,\epsilon}$ (\ref{eq:GaussianField}). In particular, the field (\ref{eq:MultifractalFieldIndicesIndpt}) converges in a $L_2$ sense when $\epsilon\to 0$, and it has same variance and covariance as the underlying Gaussian field (see annex \ref{ann:GF} for the properties of the covariance of the underlying Gaussian field). We will note the corresponding limiting process as $\bu^{ind}=\lim_{\epsilon\to 0}\bu^{ind,\epsilon}$. Similarly, the Gaussian field $\bu^g$ and the field $\bu^{ind}$ share the same second order structure functions (both longitudinal and transverse). 

\subsubsection{Fourth order structure function and Flatnesses}\label{sec:DerivFlatnesses}

We define the velocity increment $\delta_\ell u_i$ as
\begin{equation}\label{eq:VIFieldIndicesIndpt}
\delta_\ell u_i^{ind,\epsilon} = u_i^{ind,\epsilon}(\bell/2) - u_i^{ind,\epsilon}(-\bell/2) = \frac{1}{c_\epsilon}\int\Phi^{\epsilon,\bell}_{ik}(\bz)\left(\e^{\gamma \tX^\epsilon(\bz)}\right)_{kl}W_l(\diff z),
\end{equation}
where we have defined the even function
$$ \Phi^{\epsilon,\bell}_{ik}(\bx)=\phi^\epsilon_{ik}(\bx+\bell/2)-\phi^\epsilon_{ik}(\bx-\bell/2). $$ 
Of special interest is the fourth order structure function $\E (\delta_\ell u_i^\epsilon)^{4}$ (longitudinal and transverse cases), and the respective flatnesses, i.e. $\mathcal F^{\parallel}$ and $\mathcal F^{\perp}$ (\ref{eq:Flatnesses}). We get (no summation over repeated index $i$ implied),
\begin{align}\label{eq:FourthOrderSF}
\E (\delta_\ell & u_i^{\epsilon,ind})^{4} = \notag\\
&\frac{3}{c_\epsilon^{4}}\int\Phi^{\epsilon,\bell}_{ik_{1}}(\bs{z_2})\Phi^{\epsilon,\bell}_{ik_{2}}(\bs{z_2})\Phi^{\epsilon,\bell}_{ik_{3}}(\bs{z_4})\Phi^{\epsilon,\bell}_{ik_{4}}(\bs{z_4})\E \left[\left(\e^{2\gamma \tX^\epsilon(\bs{z_2})}\right)_{k_{1}k_{2}}\left(\e^{2\gamma \tX^\epsilon(\bs{z_4})}\right)_{k_{3}k_{4}}\right]\diff z_{2}\diff z_{4},
\end{align}
entering therefore the covariance of the matrix multiplicative chaos $\e^{2\gamma \tX^\epsilon}$, which is analytically derived in \cite{CheRho13}. In annex \ref{ann:IndeptCase}, we show that the fourth-order structure function converges when $\epsilon \to 0$ if we choose $\gamma^2<H/2$, and behaves asymptotically in the limit of vanishing scale $\ell\to 0$ as
\begin{align*}
\E (\delta_\ell u^{ind}_i)^{4} \build{\sim}_{\ell\to 0}^{} C_4^{ind}\frac{\ell^{4H-4\gamma^2}}{\ln\frac{1}{\ell}},
\end{align*}
where the multiplicative constant $C_4^{ind}$ is derived in annex \ref{ann:IndeptCase}. This shows that longitudinal and transverse fourth order structure functions have the same scaling behaviours. More precisely, we obtain
$$ \mathbb E (\delta_\ell^{\parallel} u^{ind})^4 \build{\sim}_{\ell\rightarrow 0}^{} C_4^{ind,\parallel}\frac{\ell^{4H-4\gamma^2}}{\ln\frac{1}{\ell}} \, \mbox{ and } \, \mathbb E (\delta_\ell^{\perp} u^{ind})^4 \build{\sim}_{\ell\rightarrow 0}^{} C_4^{ind,\perp}\frac{\ell^{ 4H-4\gamma^2}}{\ln\frac{1}{\ell}}$$
where the different constants $C_4^{ind,\parallel}$ and $C_4^{ind,\perp}$ are also given in annex \ref{ann:IndeptCase}. This shows that the velocity field $\bu^{ind}$ (\ref{eq:MultifractalFieldIndicesIndpt}), built assuming independence of $\tX$ and $\bW$, is intermittent, and the respective flatnesses (\ref{eq:Flatnesses}) behave as power-laws times a logarithmic correction with the scale $\ell$ (see annex \ref{ann:IndeptCase}):
\begin{equation}\label{eq:PredIndFlat}\mathcal F^{\parallel}(\ell)\build{\sim}_{\ell\to 0}^{} \frac{C_4^{ind,\parallel}}{(C_2^{ind,\parallel})^2} \frac{\ell^{-4\gamma^2}}{\ln\frac{1}{\ell}} \, \mbox{ and } \, \mathcal F^{\perp}(\ell)\build{\sim}_{\ell\to 0}^{} \frac{C_4^{ind,\perp}}{(C_2^{ind,\perp})^2} \frac{\ell^{-4\gamma^2}}{\ln\frac{1}{\ell}}. 
\end{equation}
Thus, according to this asymptotic prediction (\ref{eq:PredIndFlat}), the power-law exponents of the flatness are the same and related to the parameter $\gamma$, namely  $\beta^{\parallel}(\gamma)=\beta^{\perp}(\gamma) = -4\gamma^2$. 

As we can observe in figure \ref{fig:flatness_fit}, this asymptotic prediction performs poorly against our numerical results. We will see in section \ref{sec:FSSFlat} that this quantitative discrepancy can be explained while taking into account finite scale $\ell > 0$ corrections, as it is necessarily done while fitting power-laws of flatnesses estimated in numerical simulations.

\subsubsection{Heuristics for higher order structure functions}
It is easy to see, and shown in annex \ref{ann:IndeptCase}, that all odd order structure functions vanish under the hypothesis of independence: the assumption of independence prevents us from studying the physics of energy transfer. Let us study now the even higher-order structure functions and let us consider the $2n$-order moment of velocity increments $\E (\delta_\ell  u_i^{\epsilon,ind})^{2n}$, for $n\in \mathbb N$, given in equation (\ref{eq:2nIndptSF}). Doing so, we are left considering the $n^{\mathrm{th}}$-correlator of the matrix exponential of $2\gamma \tX^\epsilon$:
$$\E \left[\prod_{q=1}^{n}\left(\e^{2\gamma \tX^\epsilon(\bs{z_{2q}})}\right)_{k_{2q-1}k_{2q}}\right].$$
For $n>2$ calculations are tedious, but, based on a conjecture of \cite{CheRho13} and an appropriate range of orders $q$, we can write that the structure functions behave as
$$ \lim_{\ell \to 0}\frac{\ln  \mathbb E (\delta_\ell^{\parallel} u^{ind,\parallel,\perp})^{2q}}{\ln \ell} = \zeta_{2q}^{ind},$$
with a similar spectrum of exponents for both longitudinal and transverse structure functions given by a quadratic function of the order $q$, namely
$$ \zeta_{q}^{ind} = qH-\frac{q(q-2)}{2}\gamma^2,$$
showing that indeed, the parameter $\gamma$ fully determines the intermittent properties of the velocity field.

\subsubsection{Finite size corrections and interpretation of numerics}\label{sec:FSSFlat}

As far as flatnesses are concerned, we have found in our numerical simulations, for which results are displayed in figures \ref{fig:flatness}(b,d), a power-law exponent $\beta^{\parallel,\perp}(\gamma)$, defined in (\ref{eq:FlatnessesExpo}), much smaller (in magnitude) than our prediction $4\gamma^2$ (\ref{eq:PredIndFlat}). This is shown in figure \ref{fig:flatness_fit}(b). In this section we propose to explain this surprising fact while taking into account finite-size corrections based on our predictions before looking at the asymptotic limit $ \ell\to 0$. 

We have seen while deriving the flatnesses of velocity increments (section \ref{sec:DerivFlatnesses} and annex \ref{ann:IndeptCase}) that the covariance of the matrix chaos $\e^{2\gamma\tX^\epsilon}$ enters the expression of the fourth-order structure function, as shown in (\ref{eq:FourthOrderSF}). The mathematical theory developed in \cite{CheRho13} allows us not only to derive its asymptotic structure in the double limits $\epsilon\to 0$ and then $\ell\to 0$ as we have already seen, but also in the limit $\epsilon \to 0$ at a finite scale $\ell$. Recall that in the double limits, we have found a remaining logarithmic correction to the power-laws of the flatnesses (see equation \ref{eq:PredIndFlat}). The purpose of this section is to explore the behaviour of the matrix chaos covariance when the scale $\ell$ is finite, after taking the limit $\epsilon\to 0$, that eventually leads to intermittency with logarithmic corrections at vanishing scales $\ell$.

Clearly, the dependence on the scale $\ell$ of the flatnesses is linked to this matrix chaos covariance. Under the hypothesis of isotropy, we can show that the matrix chaos covariance depends on only two scalar functions $f(\ell)$ and $g(\ell)$ defined as
$$\frac{1}{c_\epsilon^4}\E \left[\left(\e^{2\gamma \tX^\epsilon(0)}\right)_{k_{1}k_{2}}\left(\e^{2\gamma \tX^\epsilon(\bh)}\right)_{k_{3}k_{4}}\right] \build{\rightarrow}_{\epsilon\to 0}^{} f(h)\delta_{k_1k_2}\delta_{k_3k_4} + g(h)\left[\delta_{k_1k_3}\delta_{k_2k_4}+\delta_{k_1k_4}\delta_{k_2k_3} \right]. $$
We give the expressions of $f$ and $g$ in (\ref{eq:fFiniteSize}) and (\ref{eq:gFiniteSize}), that are intermediate steps before taking the limit $\ell \to 0$. As we show in annex \ref{ann:IndeptCase}, the quantities $f$ and $g$ are responsible for the intermittent correction of the fourth order structure functions, the remaining kernels $\Phi^{\bell}$ entering the expressions of the full structure functions (\ref{eq:FourthOrderSF}) participate mainly to the scaling $4H$. When renormalizing by the square of the second-order structure function, defining thus the respective flatnesses, we focus only on the intermittent corrections. Thus, we will approximate the flatness exponent $\beta^{\parallel,\perp}(\gamma)$ by the logarithmic derivative of the contribution associated to $f$ and $g$, to write
\begin{equation}\label{eq:ApproxExpoFlat}
\frac{\partial \ln \mathcal F^{\parallel,\perp}(\ell)}{\partial \ln \ell} =  \beta^{\parallel,\perp}(\gamma) \approx \frac{\partial \ln[3f(\ell)+6g(\ell)]}{\partial \ln \ell}.
\end{equation} 
As we have seen also, a key quantity that enters the expression of $f$ and $g$ (equations \ref{eq:fFiniteSize} and \ref{eq:gFiniteSize}) is the covariance of the diagonal elements of $\tX$, i.e. $\sigma^2_{h}$ (equation \ref{eq:AsympSigmaCorrAlpha}), and we will write it as
\begin{equation}\label{eq:AsympSigmaCorrAlphaScale}
\sigma^2_{\ell} = \ln\left(\frac{L}{\ell}\right) + \alpha,
\end{equation}
where $\alpha$ is a constant independent of the scale $\ell$ (it is more generally a bounded function of the scale, but we will neglect this functional dependence). Obviously, the constant $\alpha$ is negligible in front of the logarithm (\ref{eq:AsympSigmaCorrAlphaScale}) when $\ell\to 0$. It is not the case when $\ell$ is finite. We have estimated this constant in our numerical simulation and we find $\alpha \approx -0.5$ (data not shown). Using equation (\ref{eq:ApproxExpoFlat}), we evaluate the logarithmic derivative at the scale $\ell=0.2$, using the form of the covariance given in (\ref{eq:AsympSigmaCorrAlphaScale}) with thus $\alpha = -0.5$. We display the result of this fit, as a function of $\gamma$, in figure \ref{fig:flatness_fit}(b). The comparison with numerical data is fairly good at low values of $\gamma$ and deteriorates at higher values. Several remarks can be made at this stage to justify the level of adequacy of our fit with numerical data. First, the model used to fit our data has been obtained in the limit of vanishing resolutions $\epsilon\to 0$, whereas it can remain some finite-$\epsilon$ corrections when looking at a numerical simulation. Secondly, the parameter $\alpha$ rigorously should be seen as a bounded function of the scales $\ell$. We took it as a constant for the sake of simplicity. Thirdly, relation (\ref{eq:ApproxExpoFlat}) is only an approximation, and there could be additional finite scale corrections related to the underlying Gaussian velocity field. Recall indeed, as shown in appendix \ref{ann:GF}, that the Gaussian velocity field exhibits exact power laws in the double limits $\epsilon\to 0$ and $\ell\to 0$, whereas we focus here on finite scale corrections. On the theoretical side, these corrections are difficult to obtain and will depend on the precise shape of the large-scale cut-off function $\varphi_L$ entering the definition of the velocity field (\ref{eq:MultifractalFieldIndicesIndpt}). Finally, let us note that the influence of the free parameter $\alpha$ entering equation (\ref{eq:AsympSigmaCorrAlphaScale}) is only quantitatively substantial for higher values of $\gamma$. We explain in this way the surprising fact that the estimated values of $\beta^{\parallel,\perp}(\gamma)$ based on our simulations differ from the asymptotic prediction $-4\gamma^2$ by taking into account finite-size corrections.

Under the independence assumption, we are thus able to quantify finite-size corrections to the scalings. Going back to the full vector field (\ref{eq:MultifractalField}), we observe in figure \ref{fig:flatness_fit}(a) that, likewise, longitudinal and transverse structure functions seem to be affected by finite size corrections. We furthermore observe a slight difference between the longitudinal and transverse cases: according to our numerical simulations, it seems that transverse intermittency corrections are bigger than the longitudinal ones. The underlying strong correlation between the chaos and the white noise prevents us from deriving analytically the asymptotic regime and thus, we cannot conclude at this stage whether this slight difference will remain at vanishing resolution and vanishing scale.

\section{Energy transfer: skewness phenomenon}\label{sec:EnerTrans}

Let us now turn back to the full (i.e. correlated) vector field (\ref{eq:MultifractalField}) that we recall here for convenience,
 \begin{equation}\label{eq:MultifractalFieldCorr}
u_i^\epsilon(\bx) =\int\phi_{ik}^\epsilon(\bx-\bz)\left(\e^{\gamma \tX^\epsilon(\bz)}\right)_{kl}W_l(\diff z),
\end{equation}
where
$$ \phi_{ik}^\epsilon(\bx) = -\epsilon_{ijk}\varphi_L(\bx)\frac{x_j}{|\bx|_\epsilon^{\frac{5}{2}-H}}, $$ 
and the following field (\ref{eq:Xepsilon}) is built from the very same vector white noise $\bW$ that enters the definition of the velocity field (\ref{eq:MultifractalFieldCorr}):
\begin{equation}
\tX^\epsilon(\bx) = \sqrt{\frac{15}{32\upi}}\int_{|\bx-\by|\le L} \frac{\bx-\by}{|\bx-\by|^{7/2}_\epsilon}\otimes [(\bx-\by)\wedge \bW(\diff y)]+ [(\bx-\by)\wedge \bW(\diff y)] \otimes \frac{\bx-\by}{|\bx-\by|^{7/2}_\epsilon}.\notag
\end{equation}
The same vector noise $\bW$ enters both the velocity field equation (\ref{eq:MultifractalFieldCorr}) and the matrix field implying peculiar correlations, fully given by the correlators $\Gamma^p_\epsilon$, that read, for $1\le p \le 3$, 
\begin{align}\label{eq:GammaP}
\Gamma^p_\epsilon(\bx-\by) &= \mathbb E\left[\tX^\epsilon(\bx)\frac{W_p(\diff y)}{\diff y}\right]\notag \\
&=  \sqrt{\frac{15}{32\upi}}\left[\frac{\bx-\by}{|\bx-\by|^{7/2}_\epsilon}\otimes [(\bx-\by)\wedge \bs{e_p}]+ [(\bx-\by)\wedge \bs{e_p}] \otimes \frac{\bx-\by}{|\bx-\by|^{7/2}_\epsilon}\right] \mbox{ ,}
\end{align}
where we have defined the unit vector $\bs{e_p}$ with $(\bs{e_p})_i = \delta_{ip}$. Let us also remark that at a given finite $\epsilon>0$,
$$ \Gamma^p_\epsilon(0)  =0.$$
The components of the $3\times 3$ matrices $ \Gamma^p_\epsilon$ are noted $\Gamma^p_{\epsilon,ij}$.

Obtaining the exact statistical properties of the velocity field $\bu^\epsilon$ (\ref{eq:MultifractalFieldCorr}) is a difficult task. The very peculiar correlation between $\tX^\epsilon$ and $\bW$, fully encoded in the correlator $\Gamma^p_\epsilon$ (\ref{eq:GammaP}), and the non-commutative nature of the field of matrices $\tX^\epsilon$ make the calculation out of reach at the present time. As an example, we do not know today how to perform such a calculation even for the variance of the field $\bu^\epsilon$. Instead, in order to interpret the numerical evidences of energy transfer observed in \cite{CheRob10}, we propose to do a simpler calculation, namely a perturbative expansion in power of $\gamma$ that we hope will capture several key ingredients of the physics of energy transfer. Making such an expansion prevents the analysis of intermittent corrections since, in nature, the intermittency phenomenon cannot be treated, as far as we know, with such an expansion. 

As in the independent case $\bu^{\epsilon,ind}$ (\ref{eq:MultifractalFieldIndicesIndpt}), we expect for the full vector field $\bu^{\epsilon}$ (\ref{eq:MultifractalFieldCorr}) a normalization constant such that $\bu^{\epsilon}$ is of finite variance. Recall that the elements of $\tX^\epsilon$ are Gaussian random variables whose variance diverges logarithmically with $\epsilon$ (cf. equation \ref{eq:AsympSigmaEps}). Indeed, for the independent case, we have shown that the velocity field $\bu^{\epsilon,ind}$ has to be normalized by a constant $c_\epsilon$ that itself diverges with $\epsilon$. It has been shown in \cite{CheRho13} that the multiplicative chaos $\e^{\gamma \tX^\epsilon}$ has to be renormalized in order to define a proper random variable (see also \citealt{RhoVar14} for a review on this topic). As far as $\bu^{\epsilon}$ (\ref{eq:MultifractalFieldCorr}) is concerned, since we do not know how to get the variance, we cannot make such a normalization constant explicit, but is expected to be of the order of $\E [\tr (\e^{\gamma \tX^\epsilon})]$ \citep[as it has been proved for the chaos in][]{CheRho13} or $\sqrt{\E [\tr (\e^{2\gamma \tX^\epsilon})]}$ (as we have shown for the independent case). In the following perturbative expansion, we expect then a contribution of order $\gamma^2$ (and all the following even powers of $\gamma$) from this constant. Since we do not know it explicitly, we will limit ourselves to a first order expansion in $\gamma$, which will not have any contribution from this possible unknown renormalizing constant. We will see then that such a first order expansion exhibit energy transfer.

\subsection{First order expansion of the covariance}

At a given finite $\epsilon$, the covariance of the vector field is given by
\begin{align*}
\E [u^{\epsilon}_i(0)u^{\epsilon}_p(\bh)] = \int\phi^\epsilon_{ik}(-\bs{z_1})\phi^\epsilon_{pq}(\bh-\bs{z_2})\E \left[\left(\e^{\gamma \tX^\epsilon(\bs{z_1)}}\right)_{kl}\left(\e^{\gamma \tX^\epsilon(\bs{z_2})}\right)_{qr}W_l(\diff z_1)W_r(\diff z_2)\right].
\end{align*}
Expanding the matrix exponentials up to first order gives
\begin{align*}
&\E \left[\left(\e^{\gamma \tX^\epsilon(\bs{z_1})}\right)_{kl}\left(\e^{\gamma \tX^\epsilon(\bs{z_2})}\right)_{qr}W_l(\diff z_1)W_r(\diff z_2)\right]=\E \left[W_k(\diff z_1)W_q(\diff z_2)\right]\\
+&\gamma \left( \E [X_{kl}^\epsilon(\bs{z_1})W_l(\diff z_1)W_q(\diff z_2)]+\E [X_{qr}^\epsilon(\bs{z_2})W_k(\diff z_1)W_r(\diff z_2)]\right)+o_\epsilon(\gamma),
\end{align*}
The $0^{\mathrm{th}}$-order term $\E \left[W_k(\diff z_1)W_p(\diff z_2)\right]$ gives rise to the underlying Gaussian contribution. It obviously converges when $\epsilon\to 0$. The first order term proportional to $\gamma$ vanishes, since the expectation of the product of an odd number of zero average Gaussian random variables always vanishes. We are thus left with
\begin{align*}
\E [u^{\epsilon}_i(0)u^{\epsilon}_p(\bh)] = (\phi^\epsilon_{ik}\star\phi^\epsilon_{pq})(\bh) +o_\epsilon(\gamma),
\end{align*}
which is the covariance of the underlying Gaussian velocity field (see annex \ref{ann:GF}). There remains a possible dependence on $\epsilon$ in the remaining contributions $o_\epsilon(\gamma)$. We will neglect it and assume that the following limit when $\epsilon\to 0$ makes sense:
\begin{align*}
\E [u_i(0)u_p(\bh)] = (\phi_{ik}\star\phi_{pq})(\bh) +o(\gamma).
\end{align*}
Thus, at first order in $\gamma$, the velocity field $\bu^\epsilon$ (\ref{eq:MultifractalFieldCorr}) has the same covariance as the underlying Gaussian field $\bu^{g,\epsilon}$.

\subsection{First order expansion of the third order structure function}

We recall that the velocity increment $\delta_\ell u_i$ is defined as
 \begin{equation}\label{eq:redefStructFunct}
\delta_\ell u_i^\epsilon = u_i^\epsilon(\bell/2) - u_i^\epsilon(-\bell/2) = \int\Phi^{\epsilon,\bell}_{ik}(\bz)\left(\e^{\gamma \tX^\epsilon(\bz)}\right)_{kl}W_l(\diff z),
\end{equation}
where we have defined the even function
$$ \Phi^{\epsilon,\bell}_{ik}(\bx)=\phi^\epsilon_{ik}(\bx+\bell/2)-\phi^\epsilon_{ik}(\bx-\bell/2). $$ 
Assuming no summation over the index $i$, we find
\begin{align*}
\E(\delta_\ell u_i^\epsilon)^3 &= \int\Phi^{\epsilon,\bell}_{ik_1}(\bs{z_1})\Phi^{\epsilon,\bell}_{ik_2}(\bs{z_2})\Phi^{\epsilon,\bell}_{ik_3}(\bs{z_3})\\
&\E\left[\left(\e^{\gamma \tX^\epsilon(\bs{z_1})}\right)_{k_1l_1}\left(\e^{\gamma \tX^\epsilon(\bs{z_2})}\right)_{k_2l_2}\left(\e^{\gamma \tX^\epsilon(\bs{z_3})}\right)_{k_3l_3}W_{l_1}(\diff z_1)W_{l_2}(\diff z_2)W_{l_3}(\diff z_3)\right].
\end{align*}
Performing similar expansions as for the covariance case, we obtain, up to first order in $\gamma$,
\begin{align*}
\E&\left[\left(\e^{\gamma \tX^\epsilon(\bs{z_1})}\right)_{k_1l_1}\left(\e^{\gamma \tX^\epsilon(\bs{z_2})}\right)_{k_2l_2}\left(\e^{\gamma \tX^\epsilon(\bs{z_3})}\right)_{k_3l_3}W_{l_1}(\diff z_1)W_{l_2}(\diff z_2)W_{l_3}(\diff z_3)\right]\\
&=\gamma \E\Big[ X^\epsilon_{k_1l_1}(\bs{z_1})W_{l_1}(\diff z_1)W_{k_2}(\diff z_2)W_{k_3}(\diff z_3)+X^\epsilon_{k_2l_2}(\bs{z_2})W_{k_1}(\diff z_1)W_{l_2}(\diff z_2)W_{k_3}(\diff z_3)\\
&+X^\epsilon_{k_3l_3}(\bs{z_3})W_{k_1}(\diff z_1)W_{k_2}(\diff z_2)W_{l_3}(\diff z_3)\Big]+ o_\epsilon(\gamma).
\end{align*}
The three terms on the right-hand side of the former development give similar contributions once inserted in the expression of $\E(\delta_\ell u_i^\epsilon)^3$. Focusing for example on the first term, using the fact that the product of a even number of zero-averaged Gaussian variables factorizes into pair products under an expectation value (Isserlis' theorem) and omitting for convenience the obvious dependence of $\Gamma$ on $\epsilon$, we obtain
\begin{align*}
\E\Big[ &X^\epsilon_{k_1l_1}(\bs{z_1})W_{l_1}(\diff z_1)W_{k_2}(\diff z_2)W_{k_3}(\diff z_3)\Big]\\
&=\Gamma_{k_1l_1}^{k_2}(\bs{z_1}-\bs{z_2})\E \Big[W_{l_1}(\diff z_1)W_{k_3}(\diff z_3)\Big]\diff z_2+\Gamma_{k_1l_1}^{k_3}(\bs{z_1}-\bs{z_3})\E \Big[W_{l_1}(\diff z_1)W_{k_2}(\diff z_2)\Big]\diff z_3,
\end{align*}
where we have used that $\Gamma_{k_1l_1}^{l_1}(0)=0$. Once again, the two terms on the right-hand side of the former equality give similar contributions once inserted in the expression of $\E(\delta_\ell u_i^\epsilon)^3$. We end up with, using the change of variable $\bh =\bs{z_1}-\bs{z_2}$ and performing the remaining integration over $\bs{z_1}$,
$$\E(\delta_\ell u_i^\epsilon)^3 =6\gamma \int (\Phi^{\epsilon,\bell}_{ik_1}\Phi^{\epsilon,\bell}_{il_1}\star \Phi^{\epsilon,\bell}_{ik_2})(-\bh)\Gamma_{k_1l_1}^{k_2}(\bh)\diff h + o_\epsilon(\gamma). $$
Note that because of the parity of the function $\Phi^{\epsilon,\bell}_{ik_1}$, the function $(\Phi^{\epsilon,\bell}_{ik_1}\Phi^{\epsilon,\bell}_{il_1}\star \Phi^{\epsilon,\bell}_{ik_2})(\bh)$ is even. Thus, if the correlator $\Gamma^{p}(\bh)$ was odd, which is the case when $\tX^\epsilon$ is proportional to the identity matrix (i.e. considering a scalar multiplicative chaos), the third order moment of velocity increments would have vanished. This is consistent with the conclusions of \cite{RobVar08}. In our case, the use of a matrix field $\tX^\epsilon$ ensures a non trivial third order moment. Without loss of generality, we work for instance with the first velocity component $i=1$, and we rename repeated indices as
$$\E(\delta_\ell u_1^\epsilon)^3 =6\gamma \int (\Phi^{\epsilon,\bell}_{1i}\Phi^{\epsilon,\bell}_{1j}\star \Phi^{\epsilon,\bell}_{1k})(\bh)\Gamma_{ij}^{k}(\bh)\diff h + o_\epsilon(\gamma). $$
We can write
\begin{align*} 
\Gamma^k_{ij} (\bh) =\sqrt{\frac{15}{32\upi}}\frac{h_p}{|\bh|^{7/2}_\epsilon}\left(\epsilon_{jpk}h_i+\epsilon_{ipk}h_j\right),
\end{align*}
to obtain
$$\E(\delta_\ell u_1^\epsilon)^3 =12\gamma \sqrt{\frac{15}{32\upi}}\epsilon_{ipk}\int (\Phi^{\epsilon,\bell}_{1i}\Phi^{\epsilon,\bell}_{1j}\star \Phi^{\epsilon,\bell}_{1k})(\bh)\frac{h_ph_j}{|\bh|^{7/2}_\epsilon}\diff h + o_\epsilon(\gamma). $$
Consider now a longitudinal velocity increment, define the unit-vector $\bs{e_1}$ along the first direction such that $\ell = \ell \bs{e_1}$, i.e. $(\bs{e_1})_i = \delta_{1i}$. We have
$$ \Phi^{\epsilon,\bell \bs{e_1}}_{1k}(\bx)=-\epsilon_{1jk}x_j\left[\frac{\varphi_L(\bx+\ell \bs{e_1}/2)}{|\bx+\ell \bs{e_1}/2|_\epsilon^{\frac{5}{2}-H}}-\frac{\varphi_L(\bx-\ell \bs{e_1}/2)}{|\bx-\ell \bs{e_1}/2|_\epsilon^{\frac{5}{2}-H}}\right],$$ 
and we obtain 
\begin{equation}\label{eq:DefDepsilon}
\E(\delta_\ell^{\parallel} u^\epsilon)^3 =\gamma  \mathcal D_\epsilon(\ell) \ell^{3H} + o_\epsilon(\gamma),
\end{equation}
with
$$\mathcal D_\epsilon(\ell) = 12 \ell^{-3H}\sqrt{\frac{15}{32\upi}}\epsilon_{ipk}\int (\Phi^{\epsilon,\ell \bs{e_1}}_{1i}\Phi^{\epsilon,\ell \bs{e_1}}_{1j}\star \Phi^{\epsilon,\ell \bs{e_1}}_{1k})(\bh)\frac{h_ph_j}{|\bh|^{7/2}_\epsilon}\diff h.$$
In this expression for $\mathcal D_\epsilon(\ell)$, take now the limit $\epsilon\to 0$, assuming that all the integrals converge. Noticing that
$$\Phi^{\ell \bs{e_1}}_{1k}(\ell \bx) \build{\sim}_{\ell\to 0}^{}\ell^{H-\frac{3}{2}}\tilde{\Phi}^{\bs{e_1}}_{1k}(\bx) $$
with 
$$\tilde{\Phi}^{\bs{e_1}}_{1k}(\bx) \equiv -\epsilon_{1qk}\varphi_L(0)x_q\left[\frac{1}{|\bx+\bs{e_1}/2|^{\frac{5}{2}-H}}-\frac{1}{|\bx-\bs{e_1}/2|^{\frac{5}{2}-H}}\right]=-\epsilon_{1qk}\varphi_L(0)x_q \mathcal H_H(\bx),$$
we obtain
\begin{align*}
\mathcal D(\ell) = \lim_{\epsilon\to 0}\mathcal D_\epsilon(\ell) &= 12 \ell^{-3H}\sqrt{\frac{15}{32\upi}}\epsilon_{ipk}\int (\Phi^{\ell \bs{e_1}}_{1i}\Phi^{\ell \bs{e_1}}_{1j}\star \Phi^{\ell \bs{e_1}}_{1k})(\bh)\frac{h_ph_j}{|\bh|^{7/2}}\diff h\\
&\build{\sim}_{\ell\to 0}^{}12 \sqrt{\frac{15}{32\upi}}\epsilon_{ipk}\int (\tilde{\Phi}^{\bs{e_1}}_{1i}\tilde{\Phi}^{\bs{e_1}}_{1j}\star \tilde{\Phi}^{\bs{e_1}}_{1k})(\bh)\frac{h_ph_j}{|\bh|^{7/2}}\diff h,
\end{align*}
which shows that
\begin{align}\label{eq:PredD}
D&= \lim_{\ell\to 0}\mathcal D(\ell) =\mathcal D(0)  \notag\\ 
&=-12 \sqrt{\frac{15}{32\upi}}\varphi_L^3(0)\epsilon_{ipk}\epsilon_{1qi}\epsilon_{1lj}\epsilon_{1mk}\int z_qz_l(z_m+h_m)\mathcal H_H^2(\bz)\mathcal H_H(\bz+\bh)\frac{h_ph_j}{|\bh|^{7/2}}\diff z\diff h\notag \\
&=12 \sqrt{\frac{15}{32\upi}}\varphi_L^3(0)\epsilon_{1qm}\epsilon_{1lj}\int z_qz_l(z_m+h_m)\mathcal H_H^2(\bz)\mathcal H_H(\bz+\bh)\frac{h_1h_j}{|\bh|^{7/2}}\diff z\diff h\notag \\
&= 12\sqrt{\frac{15}{32\upi}}\varphi_L^3(0)\int \mathcal H_H^2(\bz)\mathcal H_H(\bz+\bh)\frac{h_1[(\bz\wedge \bh)_1]^2}{|\bh|^{7/2}}\diff z\diff h.
\end{align}
This shows that up to first order in $\gamma$, the third order moment of velocity increments (\ref{eq:DefDepsilon}), if the term $o_\epsilon(\gamma)$ remains bounded when $\epsilon \to 0$, behaves as
\begin{align}\label{eq:PredFinal3rdMom}
 \E(\delta_\ell^{\parallel} u)^3 \build{\sim}_{\ell\to 0}^{}\gamma  D\ell^{3H} + o(\gamma),
 \end{align}
where the constant $D$ given in (\ref{eq:PredD}) is not trivially zero.

\subsubsection{Numerical study of the skewness}

\begin{figure}
\begin{center}
\epsfig{file=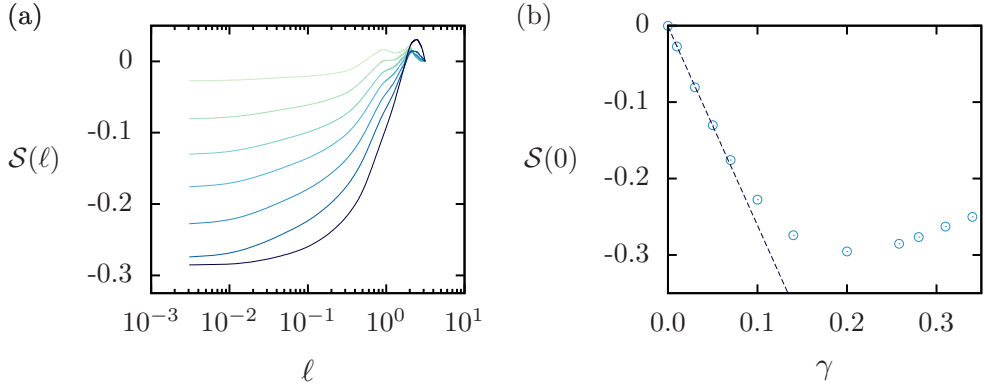,width=13cm}
\end{center}
\caption{\label{fig:Skew} (a) Skewness of longitudinal velocity increments $\mathcal S(\ell)$ (equation \ref{eq:DefSkew}) as a function of the scale $\ell$, estimated from the same type of numerical simulations as the ones used in figure \ref{fig:flatness}, for $\gamma^2= 0.0001, 0.0009, 0.0025, 0.005, 0.01, 0.02, 0.067$. (b) Value of the skewness at the origin $\mathcal S(0)$ as a function of $\gamma$. We superimpose a straight line to show the linear behaviour of $\mathcal S(0)$ at small $\gamma$, as predicted by equation (\ref{eq:PredSkewness}).
}
\end{figure}

We use the same kind of simulations as those displayed in figure \ref{fig:flatness} to estimate the numerical value of this constant $D$ (\ref{eq:PredD}). Recall that the explicit value of the constant $D$ (\ref{eq:PredD}) involves a six-dimensional integral which is tricky to estimate through numerical integration. Instead, we use the randomness of the velocity field in order to get an estimation of it. More precisely, in order to get rid of the non-universal factor $\varphi_L(0)$, we show in figure \ref{fig:Skew} the results for the skewness $\mathcal S(\ell)$, i.e. the non dimensionalized ratio of the third order moment and the power $3/2$ of the second order moment, namely 
\begin{equation}\label{eq:DefSkew} 
\mathcal S(\ell) = \frac{\E(\delta_\ell u_1)^3}{[\E(\delta_\ell u_1)^{2}]^{3/2}}.
\end{equation}
According to the former $\gamma$-expansion (\ref{eq:PredFinal3rdMom}), we predict, at small $\gamma$, a skewness $\mathcal S(\ell)=\mathcal S(0)$ independent of the scale $\ell$, and we get
\begin{align}\label{eq:PredSkewness}
\mathcal S(0) &= \frac{\gamma D}{[C_2^{g,\parallel}]^{3/2}} + o(\gamma),
\end{align}
where the constant $C_2^{g,\parallel}$ is given in annex \ref{ann:GF} (\ref{eq:C2gparr}). We remark that, at this stage, the value of $\mathcal S(0)$ at first order in $\gamma$ is universal, in the sense that it does not depend on the large-scale quantity $\varphi_L(0)$ and on the precise underlying regularization model given in (\ref{eq:RegulNorm}).

In figure \ref{fig:Skew}(a) we display the numerical estimation of the skewness obtained from simulations of the velocity field $\bu^\epsilon$ (\ref{eq:MultifractalField}). We indeed see that, for small values of $\gamma$, the skewness $\mathcal S$ is independent of the scale $\ell$. Remark also that the skewness is negative for any $\gamma$, as it is observed in turbulence. We gather in  figure \ref{fig:Skew}(b) values of the skewness at vanishing scale $\mathcal S(0)$. We observe a linear dependence of $\mathcal S$ on $\gamma$, which shows (numerically) that at first order in $\gamma$, the proposed velocity field $\bu^\epsilon$ (\ref{eq:MultifractalField}) gives a non trivial, non vanishing and strictly negative skewness. Fitting the linear behaviour of $\mathcal S(0)$ at small $\gamma$, we find $\mathcal S(0)\approx -2.6\gamma + o(\gamma)$, showing that, at first order in $\gamma$, our prediction (\ref{eq:PredSkewness}) makes sense. As far as turbulence is concerned, we have found in section \ref{sec:IntermittencyNS} that the particular value $\gamma^2=0.067$ gives a fairly realistic behaviour of the flatness when compared against empirical findings. When inserted into our first order prediction (\ref{eq:PredSkewness}), we obtain $\mathcal S(0)\approx -0.67$, which overestimates typical values obtained in experiments (see \citealt{Fri95} and \citealt{CheCas12}). Let us comment on this discrepancy. First of all, real turbulence exhibits intermittent corrections, making the skewness (equation \ref{eq:DefSkew}) slightly dependent on the scale $\ell$ and a skewness of derivatives $\mathcal S(0)$ dependent on the Reynolds number (or on $\epsilon$ in the present picture). These corrections cannot be seen with such a perturbative expansion as proposed in equation (\ref{eq:PredSkewness}). Furthermore, the typical intermittency parameter $\gamma^2=0.067$, or equivalently $\gamma=0.26$ is out of the range of observed linear behaviour of $\mathcal S(0)$ with $\gamma$ (figure \ref{fig:Skew}b). These two facts can explain this discrepancy. If now we take a look at the value of the skewness we are obtaining in the numerical simulation for the parameter $\gamma=0.26$ , without invoking the perturbative expansion (see figure \ref{fig:Skew}b), we find $\mathcal S(0)\approx -0.28$ which is very close to what is obtained in real flows \citep[see][]{CheCas12}. This theoretical and numerical study shows that indeed the proposed velocity field $\bu^\epsilon$ (\ref{eq:MultifractalField}), and for the first time as far as we know, exhibits non vanishing and realistic energy transfer.

\section{Conclusion and perspectives}\label{sec:Conclusion}

We have studied the statistical properties of an explicit random velocity field (\ref{eq:MultifractalField}) able to reproduce the main properties of a fully developed turbulent flow, as observed in experiments and numerical simulations. To do so, we have performed simulations up to $2048^3$ grid points, and developed analytical techniques when calculations were possible. This claimed realistic picture of homogeneous and isotropic turbulence includes the teardrop shape of the joint density of the invariants of the velocity gradient tensor and the preferential alignment of vorticity with the eigenframe of the deformation matrix (section \ref{sec:NumA}). Furthermore, assuming independence of the building block of the intermittency phenomenon, i.e. the matrix multiplicative chaos ($\e^{\gamma \tX^\epsilon}$), on the underlying Gaussian white measure ($\bW$), we are able to derive the spectrum of exponents $\zeta_q = qH-q(q-2)\gamma^2/2$ (see section \ref{sec:IndptCase}), which is thus found to be a non-linear function of the order $q$ and gives full meaning to the free parameter $\gamma$. As we explained, such a strong simplification forbids energy transfer. In order to obtain some insights in the physics of energy transfer, we perform a perturbative expansion of the full vector field in powers of $\gamma$ and show that at first order, the field exhibits a non vanishing third order moment of longitudinal velocity increments, establishing the dissipative nature of the field. This behaviour is consistent with Kolmogorov's $4/5$-law, although the precise value of the prefactor $4/5$ is not predicted since it requires a link (and thus an additional free parameter) between the regularization scale $\epsilon$ and the kinematic viscosity $\nu$. Nonetheless, we are led to the conclusion that the proposed velocity field is realistic of turbulence when we set $\gamma^2=0.067$, a value that has been obtained when comparing the power-law behaviour of the flatness (see equation \ref{eq:FlatnessesExpo}) with experimental findings, taking into account finite-size corrections to the scalings.

As far as we know, the velocity field $\bu^\epsilon$  (\ref{eq:MultifractalField}) is the first stochastic process proposed in the literature that is able to predict a non vanishing mean energy transfer across scales. At least numerically, and supported by the perturbative expansion performed in section \ref{sec:EnerTrans}, it seems that the non vanishing nature of the third order moment of longitudinal velocity increments remains in the asymptotic limit $\epsilon\to 0$. This justifies the term \textit{dissipative} used in the title of the present article and gives hope to define one day rigorously the limiting velocity field $\bu =\lim_{\epsilon\to 0}\bu^\epsilon$ as a stochastic representation of weak solutions of the Euler equations, as they are depicted in Onsager's contribution to turbulence (see the review article \citealt{EyiSre06} on this subject). In particular, the modern view of energy transfer of \cite{DucRob00} shows that standard local energy budget is possibly violated by an additional dissipative term, independent of viscosity, related to the non differential nature of the velocity field. It would be interesting to relate this view to the statistical properties of the proposed field $\bu$. To do so, new mathematical techniques are needed in order to handle the tricky correlated nature of the matrix $\tX^\epsilon$ and vector $\bW$ fields.

As perspectives, let us mention first the usefulness of such a random field in the context of mean square estimation \citep{Pap91} and related conditional averages as far as turbulent applications are concerned \citep[see for example][]{AdrMoi88}. Indeed, such a procedure can be used to provide new types of closures of the subgrid stress tensor while performing a large eddy simulation \citep{LanMos99}. More recently, it was shown that a Gaussian velocity field is able to reproduce non trivial properties of the pressure Hessian that enters the dynamics of the velocity gradient tensor \citep[see][]{Men11}. More precisely, it is shown numerically in \cite{CheLev11} that the following average of the Hessian of the pressure $p$ conditioned on the local velocity gradient tensor $\tA$, i.e. 
$$ \E\left[ \frac{\partial^2 p(\bx)}{\partial x_i\partial x_j}\Big| \tA(\bx)\right],$$ 
is well approximated, when compared against direct numerical simulations of the Navier-Stokes equations, if one assumes the velocity field to be Gaussian, such as $\bu^g$ (\ref{eq:GaussianField}). Further analytical work in this direction by \cite{WilMen14} shows that indeed the prediction of this conditional average starting from a Gaussian velocity field is able, when slightly modified, to regularize the finite-time divergence implied by the self-stretching term. These results could then be extended while assuming for the velocity field a stochastic structure such as the one we are proposing (\ref{eq:MultifractalField}). A first step in this direction could be reached assuming furthermore independence of the multiplicative chaos on the underlying white measure because this simplification makes calculations tractable (section \ref{sec:IndptCase}).

Finally let us comment on the remaining free parameter $\gamma$ that governs both energy transfer and the intermittency phenomenon in our velocity field $\bu^\epsilon$ (\ref{eq:MultifractalField}). It would be much welcome to use further constraints from the equations of motion in order to set its value that would compare in an appropriate way against experiments. Constraints on structure functions of order higher than the third one have been derived by \cite{Hil01} and \cite{Yak01}. They involve the pressure field $p$ which is fully determined by the corresponding velocity field through the Poisson equation. We could inquire if these constraints, which include pressure gradient increments, can be used to give a precise range of possible values for the intermittency coefficient $\gamma$. We keep these perspectives for future investigations.

We thank B. Castaing, K. Gawedzki and E. L\'ev\^{e}que for fruitful discussions. Warm acknowledgments to R. Robert, R. Rhodes and V. Vargas for their help at the early stage of this work. This work was supported by the ANR Liouville project, grant ANR-15-CE40-0013 of the French Agence Nationale de la Recherche. One of the authors (R.M.P.) also thanks CAPES for financial support through the scholarship process number 9497/13-7.
We gratefully acknowledge support from the PSMN (P\^{o}le Scientifique de Mod\'elisation Num\'erique) computing center of ENS de Lyon.

 \appendix

\section{Covariance structure of the field of matrices}
\label{ann:CorrX}
\subsection{General description}\label{sec:GeneCovMat}

The homogeneous field of matrices as defined in equation \ref{eq:Xepsilon}  takes on the following structure:
\begin{itemize}
\item the diagonal entries $(X^\epsilon_{11},X^\epsilon_{22},X^\epsilon_{33})$ are independent of the off-diagonal entries $(X^\epsilon_{ij})$ for $1\le i<j\le 3$,
\item the covariance matrix of the diagonal entries $\E [X^\epsilon_{ii}(0)X^\epsilon_{jj}(0)] $, for $1\le i,j\le 3$, is given by $(3/2)\sigma_\epsilon^2 \mathsfbi{I}-\frac{1}{2}\sigma_\epsilon ^2\mathsfbi{P_3}$ where $\sigma_\epsilon^2=\mathbb E[(X_{11}^\epsilon)^2]$, $\mathsfbi{I}$ the  $3\times 3$ identity matrix and $\mathsfbi{P_3}=(1)_{1\leq i,j\leq 3}$ stands for the $3\times 3$ matrix filled with the coefficient $1$ in all entries. As $\epsilon$ gets smaller, we get the following asymptotic structure:
\begin{equation}  \label{eq:AsympGammaBeta}
\sigma_\epsilon^2 \build{\sim}_{\epsilon\rightarrow 0}^{} \ln\frac{L}{\epsilon}\mbox{ .}\end{equation}

\item the off-diagonal entries $(X_{ij})_{i<j}$ are mutually independent with variance $(3/4)\sigma_\epsilon^2$.
\end{itemize}
This covariance structure is peculiar to isotropic matrices, as it is demonstrated in \cite{CheRho13} where it corresponds to the particular case of trace-free matrices. Let us now turn to the characterization of the Gaussian field of matrices (\ref{eq:Xepsilon}). It is enough, since it is Gaussian, to give the covariance structure. Up to an additive independent Gaussian matrix with a bounded covariance structure as $\epsilon \rightarrow 0$, we obtain the following asymptotic covariance of the field $\tX^\epsilon$:
\begin{itemize}
\item the diagonal entries $(X^\epsilon_{11}(\bx),X^\epsilon_{22}(\bx),X^\epsilon_{33}(\bx))_{\bx\in\R^3}$ are independent of the off-diagonal entries $((X^\epsilon_{ij}(\bx))_{1\le i<j\le 3})_{\bx\in\R^3}$,
\item When $\epsilon\rightarrow 0$, the $3\times 3$ covariance matrix of the diagonal entries $\E [X^\epsilon_{ii}(\bx)X^\epsilon_{jj}(\by)] $, for $1\le i,j\le 3$, is given by $\sigma^2_{|\bx-\by|}\Big((3/2)\mathsfbi{I}-\frac{1}{2} \mathsfbi{P_3}\Big)$, where
\begin{equation}\label{eq:AsympGammaCorrBeta}
\sigma_{|\bx-\by|}^2 = \lim_{\epsilon\to 0}\E[X^\epsilon_{11}(\bx)X^\epsilon_{11}(\by)] \build{\sim}_{|\bx-\by|\rightarrow 0}^{}\ln\frac{L}{|\bx-\by|} \mbox{ .}
\end{equation}
\item the off-diagonal entries $((X^\epsilon_{ij}(\bx))_{i<j})_{\bx\in\R^3}$ are mutually independent, each of which with covariance given by, for $i\ne j$,
$$ \E[X^\epsilon_{ij}(\bx)X^\epsilon_{ij}(\by)] \build{\sim}_{\epsilon\rightarrow 0}^{} \frac{3}{4} \sigma_{|\bx-\by|}^2 \mbox{ .}$$
\end{itemize}

\subsection{Proofs}

The field of isotropic matrices $\tX^\epsilon(\bx)$ is given by
\begin{equation}\label{eq:FieldIsoMatrix}
\tX^\epsilon(\bx) = \sqrt{\frac{15}{32\upi}}\int_{|\bx-\by|\le L} \frac{\bx-\by}{|\bx-\by|^{7/2}_\epsilon}\otimes [(\bx-\by)\wedge \bW(\diff y)]+ [(\bx-\by)\wedge \bW(\diff y)] \otimes \frac{\bx-\by}{|\bx-\by|^{7/2}_\epsilon}.
\end{equation}
Next, we are showing that indeed this explicit field leads asymptotically ($\epsilon\rightarrow 0$) to the structure previously described in section \ref{sec:GeneCovMat}.

\subsubsection{Diagonal elements}

Let us consider first the diagonal components $(X^\epsilon_{ii}(\bx))_{1\le i \le 3}$ of the random matrix-valued field (\ref{eq:FieldIsoMatrix}). For instance, consider the $X^\epsilon_{11}(\bh)$ at the spatial location $\bh$:
$$X^\epsilon_{11}(\bh) = 2\sqrt{\frac{15}{32\upi}}\int\frac{h_1-y_1}{|\bh-\by|^{7/2}_\epsilon}[(h_2-y_2)W_3(\diff y)-(h_3-y_3)W_2(\diff y)]. $$
We get
\begin{align*} \sigma_{\epsilon,h}^2=\E [X^\epsilon_{11}(\bh) X^\epsilon_{11}(0)]=  \frac{15}{8\upi}\int \frac{(h_1-y_1)y_1}{|\bh-\by|^{7/2}_\epsilon|\by|^{7/2}_\epsilon}[(h_2-y_2)y_2+(h_3-y_3)y_3]\diff y.
\end{align*}
Taking $h=0$ in the previous integral, we find, using a spherical integration and the explicit form of the regularized norm (\ref{eq:RegulNorm})
\begin{align*} \sigma_\epsilon^2=\E [(X^\epsilon_{11})^2]&=  \frac{15}{4\upi}\int \frac{y_1^2y_2^2}{|\by|^{7}_\epsilon}\diff y \\
&= \frac{15}{4\upi}\int_{\theta=0}^{\upi} \int_{\varphi=0}^{2\upi} \cos^2(\theta)\sin^3(\theta)\sin^2(\varphi)d\theta d\varphi \int_{0}^L \frac{\rho^6d\rho}{\left[\rho^2+\epsilon^2\right]^{7/2}}\\
&=  \int_{0}^L \frac{\rho^6d\rho}{\left[\rho^2+\epsilon^2\right]^{7/2}}\\
&\build{\sim}_{\epsilon\to 0}^{}\ln\frac{L}{\epsilon},
\end{align*}
which entails equation \ref{eq:AsympGammaBeta}. The asymptotic logarithmic behaviour can be easily obtained while performing the change of variable $\rho=\epsilon r$. Let us remark that the asymptotic variance of the diagonal elements of the matrix $\tX^\epsilon$ is independent of the precise regularization procedure.

Take now $h\ne 0$. Every integral converges when $\epsilon \to 0$ and we obtain 
\begin{align*}
\sigma_{h}^2=\lim_{\epsilon\to 0}\E [X^\epsilon_{11}(\bh) X^\epsilon_{11}(0)]=\frac{15}{8\upi}\int \frac{(h_1-y_1)y_1}{|\bh-\by|^{7/2}|\by|^{7/2}}[(h_2-y_2)y_2+(h_3-y_3)y_3]\diff y.
\end{align*}
We will now compute the equivalent of the former correlation function as $h\to 0$. Performing the change of variable $h\bz=\by$, i.e. $h^3\diff z=\diff y$, we find
\begin{align*}
\sigma_{h}^2=\frac{15}{8\upi}\int \frac{\left( \frac{h_1}{|\bh|}-z_1\right) z_1}{\left| \frac{\bh}{|\bh|}-\bz\right|^{7/2}|\bz|^{7/2}}\left[\left(\frac{h_2}{|\bh|}-z_2\right)z_2+\left(\frac{h_3}{|\bh|}-z_3\right)z_3\right]\diff z 
\end{align*}
where the integration domain is over $|\bz|\le L/h$. As $h\to 0$, we can always choose a constant $C$ such that for $C\le |z|\le L/h$, we have $\left|\frac{\bh}{|\bh|}-\bz\right|\approx |\bz|$. In the previous integrals, the contribution resulting from the integration over the finite domain $|\bz|\le C$ gives a bounded function of the norm $h$. The remaining contributions diverge with $h\to 0$. Thus, we get
\begin{align*}
\sigma_{h}^2&\build{\sim}_{h\to 0}^{}\frac{15}{8\upi}\int \frac{\left(\frac{h_1}{|\bh|}-z_1\right)z_1}{|\bz|^{7}}\left[\left(\frac{h_2}{|\bh|}-z_2\right)z_2+\left(\frac{h_3}{|\bh|}-z_3\right)z_3\right]\diff z= \frac{15}{8\upi}\int_{C\le |z|\le L/h} \frac{z_1^2z_2^2}{|\bz|^{7}}\diff z \\
&\build{\sim}_{h\to 0}^{} \ln\frac{L}{h},
\end{align*}
which entails equation \ref{eq:AsympGammaCorrBeta}.

Of great importance is also the cross-covariance of diagonal components, say $\E [X^\epsilon_{11}(0) X^\epsilon_{22}(\bh)]$. Recall that
$$X^\epsilon_{11}(\bh) = 2\sqrt{\frac{15}{32\upi}}\int\frac{h_1-y_1}{|\bh-\by|^{7/2}_\epsilon}[(h_2-y_2)W_3(\diff y)-(h_3-y_3)W_2(\diff y)] $$
and
$$X^\epsilon_{22}(\bh) = 2\sqrt{\frac{15}{32\upi}}\int\frac{h_2-y_2}{|\bh-\by|^{7/2}_\epsilon}[(h_3-y_3)W_1(\diff y)-(h_1-y_1)W_3(\diff y)] . $$
In the same manner as followed to compute the correlation of a diagonal component, we get 
\begin{align*} 
\E [X^\epsilon_{11}(0) X^\epsilon_{22}(\bh)]= -\frac{15}{8\upi}\int \frac{(h_1-y_1)(h_2-y_2)y_1y_2}{|\bh-\by|^{7/2}_\epsilon|\by|^{7/2}_\epsilon}\diff y,
\end{align*}
which leads to
\begin{align*} \E [X^\epsilon_{11}(0)X^\epsilon_{22}(0)]&= -\frac{15}{8\upi}\int \frac{y_1^2y_2^2}{|\by|^{7}_\epsilon}\diff y \build{\sim}_{\epsilon\to 0}^{}-\frac{1}{2}\ln\frac{L}{\epsilon}.
\end{align*}
In the same spirit, the cross-covariance of the diagonal element is given by
$$
\lim_{\epsilon\rightarrow 0}\E [X^\epsilon_{11}(0)X^\epsilon_{22}(\bh)]= \E [X_{11}(0)X_{22}(\bh)] \build{\sim}_{h|\to 0}^{} -\frac{1}{2}\ln\frac{L}{h}.$$

\subsubsection{Off-diagonal elements}

Consider now an off-diagonal element such as $X_{12}^\epsilon$. We have
\begin{align*}
X^\epsilon_{12}(\bh) = \sqrt{\frac{15}{32\upi}}\int\frac{1}{|\bh-\by|^{7/2}_\epsilon}&\left[(h_1-y_1)\Big(-(h_1-y_1)W_3(\diff y)+(h_3-y_3)W_1(\diff y)\Big)\right.\\
&\left.+\Big((h_2-y_2)W_3(\diff y)-(h_3-y_3)W_2(\diff y)\Big)(h_2-y_2)\right] . 
\end{align*}
In particular
\begin{align*}
X^\epsilon_{12}(0) = \sqrt{\frac{15}{32\upi}}\int\frac{1}{|\by|^{7/2}_\epsilon}\left[-y_1\Big(y_1W_3(\diff y)-y_3W_1(\diff y)\Big)-\Big(-y_2W_3(\diff y)+y_3W_2(\diff y)\Big)y_2\right], 
\end{align*}
such that
\begin{align*}
\E [(X^\epsilon_{12})^2] &= \frac{15}{32\upi}\int\frac{1}{|\by|^{7}_\epsilon}\left[y_1^4+y_2^4+y_1^2y_3^2+y_2^2y_3^2-2y_1^2y_2^2\right]\diff y\\
&= \frac{15}{16\upi}\int\frac{y_1^4}{|\by|^{7}_\epsilon}dy= \frac{15}{8}\int_0^{\upi}\cos^4(\theta)\sin(\theta)d\theta\int_0^{L}\frac{\rho^6}{[\rho^2+\epsilon^2]^{7/2}}d\rho\\
&= \frac{3}{4}\int_0^{L}\frac{\rho^6}{[\rho^2+\epsilon^2]^{7/2}}d\rho\\
&\build{\sim}_{\epsilon\to 0}^{} \frac{3}{4}\ln\frac{L}{\epsilon}.
\end{align*}
In the same spirit, the cross-covariance of the off-diagonal element is given by
$$
\lim_{\epsilon\rightarrow 0}\E [X^\epsilon_{12}(0)X^\epsilon_{12}(\bh)]= \E [X_{12}(0)X_{12}(\bh)] \build{\sim}_{h\to 0}^{} \frac{3}{4}\ln\frac{L}{h}.$$
It can furthermore be shown that the off-diagonal elements are mutually independent, for example
$$\E [X^\epsilon_{12}(0)X^\epsilon_{13}(0)]=0,$$ 
independent of the diagonal elements, i.e.
$$\E [X^\epsilon_{11}(0)X^\epsilon_{12}(0)]=0,$$
and the respective cross-correlation functions, i.e. $\E [X_{12}(0)X_{13}(\bh)]$ and $\E [X_{11}(0)X_{12}(\bh)]$ are bounded functions of their argument $\bh$, in particular they do not diverge logarithmically with $h$.

\section{The underlying Gaussian velocity field}\label{ann:GF}

Consider the non intermittent case (i.e. $\gamma=0$). We are left considering the following Gaussian isotropic and homogeneous incompressible velocity field $(u^{\epsilon,g}_i)_{1\le i \le 3}$ (\ref{eq:GaussianField}):
\begin{equation}\label{eq:GaussianFieldIndices}
u_i^{\epsilon,g}(\bx) = \int\phi^{\epsilon}_{ik}(\bx-\bz)W_k(\diff z),
\end{equation}
where the kernel $\phi^{\epsilon}_{ik}$ is given by
$$ \phi^{\epsilon}_{ik}(\bx) =-\epsilon_{ijk}\varphi_L(\bx)\frac{x_j}{|\bx|_{\epsilon}^{\frac{5}{2}-H}}. $$ 
\subsection{Covariance and variance}

It is easily seen that the covariance of this field is (considering only the covariance between the null vector and $\bh$ by homogeneity)
\begin{equation}\label{eq:CovGF}
\E [u^{\epsilon,g}_i(0)u^{\epsilon,g}_p(\bh)]=(\phi^{\epsilon}_{ik}\star\phi^{\epsilon}_{pk})(\bh) = \int \phi^{\epsilon}_{ik}(\bx)\phi^{\epsilon}_{pk}(\bx+\bh) \diff x.
\end{equation}
which defines the correlation product $\star$. The kernel $ \phi^{\epsilon}_{ik}(\bx)$ is singular for $\bx=0$ when $\epsilon\to 0$. Thus, the limiting integral exists if the strongest singularity obtained while considering the variance (i.e. $\bh=0$) is integrable in three dimensions. This singularity $1/|\bx|^{2({\frac{5}{2}-H}-1)}$ is integrable in three dimensions for $H>0$. In particular, for this range of Hurst exponent $H>0$, the variance of the velocity field converges and we can write
\begin{align*} 
\lim_{\epsilon\to 0}\E |\bu^{\epsilon,g}|^2 =  \E |\bu^{g}|^2 &= (\phi_{ik}\star\phi_{ik})(0)\\ 
&= \epsilon_{ijk}\epsilon_{ipk}\int \varphi_L^2(\bx)\frac{x_jx_p}{|\bx|^{5-2H}}\diff x \\
&= 2\int \varphi_L^2(\bx)\frac{|\bx|^2}{|\bx|^{5-2H}}\diff x\\ 
&= 8\upi\int_{0}^{+\infty} \varphi_L^2(\rho)\rho^{2H-1}\diff\rho, 
\end{align*}
since we have assumed that $\varphi_L$ is a radially symmetric function. Remark also that the rapid decay of the cut-off function $\varphi_L$ ensures a finite variance of the process. In the sequel, we will drop the dependence on $\epsilon$ and consider the limiting process $\bu^g$. 

More generally, as expected from an isotropic velocity field and following the notations of \cite{Bat53}, the covariance structure of this Gaussian velocity field can be written as
\begin{align*}
\E [\bu^{g}_i(0)\bu^{g}_p(\bh)]&= \delta_{ip}\int \varphi_L(\bx)\varphi_L(\bx+\bh)\frac{\bx\cdot(\bx+\bh)}{|\bx|^{\frac{5}{2}-H}|\bx+\bh|^{\frac{5}{2}-H}}\diff x\\
&-\int \varphi_L(\bx)\varphi_L(\bx+\bh)\frac{x_p(x_i+h_i)}{|\bx|^{\frac{5}{2}-H}|\bx+\bh|^{\frac{5}{2}-H}}\diff x\\
&= F(|\bh|)h_ih_p+G(|\bh|)\delta_{ip},
\end{align*}
where the functions $F$ and $G$ are radially symmetric functions.

At this stage, as it is usually done in turbulence literature \citep{Bat53}, it is convenient to decompose the covariance of the field in terms of the \textit{longitudinal} and \textit{transverse} velocity correlations defined as
$$R^g_{\parallel}(\bh) = \E [u^{g}_{\parallel}(0)u^{g}_{\parallel}(\bh)] \mbox{ and } R^g_{\perp}(\bh) = \E [u^{g}_{\perp}(0)u^{g}_{\perp}(\bh)]  $$
where $u^{g}_{\parallel}$ and $u^{g}_{\perp}$ denote velocity components parallel and normal respectively to the vector separation $\bh$. In homogeneous and isotropic turbulence, the correlations $R^g_{\parallel}(\bh)$ and $R^g_{\perp}(\bh)$ are functions of the norm $h$ only, and $R^g_{\parallel}(0)=R^g_{\perp}(0) = \frac{1}{3}\E |\bu^{g}|^2$. Furthermore we have the following decomposition:
\begin{equation}\label{eq:DecompCovGauss}
R_{ip}^g(\bh)=\E [u^{g}_i(0)u^{g}_p(\bh)]=(\phi_{ik}\star\phi_{pk})(\bh)  = \frac{R^g_{\parallel}(h)-R^g_{\perp}(h)}{h^2}h_ih_p+R^g_{\perp}(h)\delta_{ip}.
\end{equation}
The incompressible condition, i.e. $\partial R_{ip}^g(\bh)/\partial h_p=0$, finally implies that the velocity covariance  $R_{ip}^g(\bh)$ depends on a single scalar function since
$$R^g_{\perp}(h) =R^g_{\parallel}(h)+\frac{h}{2}\frac{\diff R^g_{\parallel}(h)}{\diff h} .$$

\subsection{Structure functions}

The velocity increment $\delta_\ell u^g_i$ reads
 \begin{equation}\label{eq:VIGaussianFieldIndices}
\delta_\ell u_i^g(x) = u_i^g(\bx+\bell/2) - u_i^g(\bx-\bell/2) = \int\Phi^\ell_{ik}(\bx-\bz)W_k(\diff z),
\end{equation}
where we have defined
\begin{align*} 
\Phi^{\bell}_{ik}(\bx)&=\phi_{ik}(\bx+\bell/2)-\phi_{ik}(\bx-\bell/2)\\
&=-\epsilon_{ijk}\left[\varphi_L(\bx+\bell/2)\frac{x_j+\ell_j/2}{|\bx+\bell/2|_{\epsilon}^{\frac{5}{2}-H}}+\varphi_L(\bell/2-\bx)\frac{\ell_j/2-x_j}{|\bell/2-\bx|_{\epsilon}^{\frac{5}{2}-H}}\right].
\end{align*}
Note that so defined, the kernel $\Phi^{\bell}_{ik}(\bx)$ is an even function of its argument (recall that $\varphi_L$ is a radially symmetric function), i.e.
$$\Phi^{\bell}_{ik}(\bx)=\Phi^{\bell}_{ik}(-\bx).$$ 
Also, in order to simplify the following expressions, since the field is homogeneous, we will only consider velocity increments at the position $x=0$ and note
$$\delta_\ell u_i^g \equiv \delta_\ell u_i^g(0) = u_i^g(\bell/2) - u_i^g(-\bell/2) = \int\Phi^{\bell}_{ik}(\bz)W_k(\diff z).$$

Without loss of generality (by isotropy), consider only the first component $u_1^g$ of the Gaussian vector field $\bu^g$. We get:
\begin{align*}
\E(\delta_\ell u_1^g)^2 &= (\Phi^{\bell}_{1k}\star\Phi^{\bell}_{1k})(0)\\
&=\sum_{j \neq 1} \int\left ( \varphi_L(\bz+\bell/2) \frac{z_j+\ell_j/2}{|\bz+\bell/2|^{\frac{5}{2}-H}} +  \varphi_L(\bell/2-\bz)\frac{\ell_j/2 -z_j}{|\bell/2-\bz|^{\frac{5}{2}-H}} \right)^2 \,\diff z
\end{align*}
Take $\bs{e}$ as a unit vector, consider then the displacement $\bell = \ell \bs{e}$. We find, making the change of variables $\bz=\ell \by$:
\begin{align*}
\E(\delta_\ell u_1^g)^2& =\ell^{2H}\sum_{j \neq 1} \int\left ( \varphi_L[\ell(\by+\bs{e}/2)] \frac{y_j+e_j/2}{|\by+\bs{e}/2|^{\frac{5}{2}-H}} +  \varphi_L[\ell (\bs{e}/2-\by)]\frac{e_j/2-y_j}{|\bs{e}/2-\by|^{\frac{5}{2}-H}} \right)^2\, \diff y.
\end{align*}
To conclude regarding scaling behaviour, we need to discuss the remaining $\ell$ dependence in the cutoff functions $\varphi_L$. Hereafter, we will assume that $H<1$ which ensures that the integrals converge without the need of cutoff functions. Thus, for this range of parameters $H\in ]0,1[$, longitudinal and transverse velocity increments have the same scaling behaviour given by the spectrum of exponents $\zeta^{\parallel}(2) = \zeta^{\perp}(2) = 2H $. Considering for instance the unit vectors for the longitudinal case $\bs{e}=\bs{e_1}=(1,0,0)$ and $\bs{e}=\bs{e_2}=(0,1,0)$ for the transverse one, we finally obtain
$$ \mathbb E (\delta_\ell^{\parallel} u^g)^2 \build{\sim}_{\ell\rightarrow 0}^{} C_2^{g,\parallel}\ell^{ \zeta^{\parallel}(2)} \, \mbox{ and } \, \mathbb E (\delta_\ell^{\perp} u^g)^2 \build{\sim}_{\ell\rightarrow 0}^{} C_2^{g,\perp}\ell^{\zeta^{\perp}(2)} $$
with 
$$C_2^{g,\parallel} = \varphi_L(0)^2\sum_{j \neq 1}\int\left ( \frac{y_j}{|\by+\bs{e_1}/2|^{\frac{5}{2}-H}} -  \frac{y_j}{|\bs{e_1}/2-\by|^{\frac{5}{2}-H}} \right)^2\, \diff y$$ 
and
$$C_2^{g,\perp} = \varphi_L(0)^2\sum_{j \neq 1}\int\left ( \frac{y_j+\delta_{j2}/2}{|\by+\bs{e_2}/2|^{\frac{5}{2}-H}} +  \frac{\delta_{j2}/2-y_j}{|\bs{e_2}/2-\by|^{\frac{5}{2}-H}} \right)^2\,\diff y \mbox{ .}$$ 
Notice that the constants $C_2^{g,\parallel} $ and $C_2^{g,\perp}$ are non null and positive. Using the notation
\begin{equation}\label{eq:DefHH}
\mathcal H_H(\by)= \frac{1}{|\by+\bs{e_1}/2|^{\frac{5}{2}-H}} -  \frac{1}{|\bs{e_1}/2-\by|^{\frac{5}{2}-H}},
\end{equation}
we can write the constant $C_2^{g,\parallel}$ in the following convenient way
\begin{equation}\label{eq:C2gparr}
C_2^{g,\parallel} = 2\varphi_L(0)^2\int y_2^2\mathcal H_H^2(\by)\,\diff y.
\end{equation}

Since the field is Gaussian, it is straightforward to get the higher order structure functions as:
$$ \mathbb E (\delta_\ell^{\parallel} u^g)^{2q} \build{\sim}_{\ell\rightarrow 0}^{} \frac{(2q)!}{2^qq!}(C_2^{g,\parallel})^{q}\ell^{ \zeta^{\parallel}(2q)} \, \mbox{ and } \, \mathbb E (\delta_\ell^{\perp} u)^{2q} \build{\sim}_{\ell\rightarrow 0}^{} \frac{(2q)!}{2^qq!}(C_2^{g,\perp})^q\ell^{\zeta^{\perp}(2q)} $$
with
$$\zeta^{\parallel}(q)=\zeta^{\perp}(q) = qH.$$
We observe also that all odd moments vanish for a Gaussian process, i.e. for $n\in \mathbb N$
$$ \E(\delta_\ell u_1^g)^{2n+1}=0.$$

\section{Statistical properties in the independent case}\label{ann:IndeptCase}
We recall the definition of the velocity field $\bu^{\epsilon,ind}$ (\ref{eq:MultifractalFieldIndicesIndpt})
\begin{equation}\label{eq:AnnMultifractalFieldIndicesIndpt}
u_i^{\epsilon,ind}(\bx) = \frac{1}{c_\epsilon}\int\phi_{ik}^\epsilon(\bx-\bz)\left(\e^{\gamma \tX^\epsilon(\bz)}\right)_{kl}W_l(\diff z),
\end{equation}
where again
$$ \phi_{ik}^\epsilon(\bx) =-\epsilon_{ijk}\varphi_L(\bx)\frac{x_j}{|\bx|_\epsilon^{\frac{5}{2}-H}}, $$ 
and the following \textit{matrix} field:
\begin{equation}\label{eq:AnnGaussSymmMatIndpt}
\tX^\epsilon(\bx) = \sqrt{\frac{15}{32\upi}}\int_{|\bx-\by|\le L} \frac{\bx-\by}{|\bx-\by|^{7/2}_\epsilon}\otimes [(\bx-\by)\wedge \bW'(\diff y)]+ [(\bx-\by)\wedge \bW'(\diff y)] \otimes \frac{\bx-\by}{|\bx-\by|^{7/2}_\epsilon}
\end{equation}
where now the vector noise $\bW'$ is independent of the vector noise $\bW$, namely for any $(\bx,\by)\in (\mathbb R^3)^2$ and any components $k$ and $l$ we have $\E W_k(\bx)W'_l(\by)=0$. As we will see in the following, the field (\ref{eq:AnnMultifractalFieldIndicesIndpt}) needs to be renormalized in order to converge, when $\epsilon \to 0$, towards a finite-variance process. To ensure such a finite variance, we introduce a deterministic normalization constant (which depends on $\epsilon)$
\begin{equation}\label{eq:AnnRenormIndpt} 
c_\epsilon^2 = \frac{1}{3}\E\left[\tr~ \e^{2\gamma \tX^\epsilon}\right]. 
\end{equation}
This is a standard way to renormalize the multiplicative chaos \citep{RhoVar14}.

\subsection{Mean, covariance and $L_2$-convergence}

Since $\bW$ and $\tX^\epsilon$ are independent, we easily obtain
\begin{align*}
\E u_i^{ind,\epsilon}(\bx) &= \frac{1}{c_\epsilon}\int\phi_{ik}^\epsilon(\bx-\bz)\E\left[\left(\e^{\gamma \tX^\epsilon(\bz)}\right)_{kl}W_l(\diff z)\right]\\
&= \frac{1}{c_\epsilon}\int\phi_{ik}^\epsilon(\bx-\bz)\E\left[\left(\e^{\gamma \tX^\epsilon(\bz)}\right)_{kl}\right]\E\left[W_l(\diff z)\right]\\
&=0.
\end{align*}
Thus, the vector field (\ref{eq:AnnMultifractalFieldIndicesIndpt}) is of zero average, as expected for an isotropic vector field. For the covariance we get
\begin{align*}
\E [u^{ind,\epsilon}_i(0)u^{ind,\epsilon}_p(\bh)] &= \frac{1}{c_\epsilon^2}\int\phi^\epsilon_{ik}(-\bs{z_1})\phi^\epsilon_{pq}(\bh-\bs{z_2})\E \left[\left(\e^{\gamma \tX^\epsilon(\bs{z_1})}\right)_{kl}\left(\e^{\gamma \tX^\epsilon(\bs{z_2})}\right)_{qr}W_l(\diff z_1)W_r(\diff z_2)\right]\\
&= \frac{1}{c_\epsilon^2}\int\phi^\epsilon_{ik}(-\bs{z_1})\phi^\epsilon_{pq}(\bh-\bs{z_2})\E \left[\left(\e^{\gamma \tX^\epsilon(\bs{z_1})}\right)_{kl}\left(\e^{\gamma \tX^\epsilon(\bs{z_2})}\right)_{qr}\right]\E \left[W_l(\diff z_1)W_r(\diff z_2)\right]\\
&= \frac{1}{c_\epsilon^2}\int\phi^\epsilon_{ik}(-\bz)\phi^\epsilon_{pq}(\bh-\bz)\E \left[\left(\e^{2\gamma \tX^\epsilon(\bz)}\right)_{kq}\right]\diff z.
\end{align*}
The field of matrices $\tX^\epsilon$ is isotropic, thus, according to \cite{CheRho13}, we get
$$\E \left[\left(\e^{2\gamma \tX^\epsilon(\bz)}\right)_{kq}\right] = c_\epsilon^2 \delta_{kq}, $$
where the renormalizing constant is given in (\ref{eq:AnnRenormIndpt}). We can see that the covariance structure of the vector field (\ref{eq:AnnMultifractalFieldIndicesIndpt}) is the same as the one obtained from the underlying Gaussian field (\ref{eq:GaussianField}), namely
\begin{align*}
\E [u^{ind,\epsilon}_i(0)u^{ind,\epsilon}_p(\bh)] = \E [u^{g,\epsilon}_i(0)u^{g,\epsilon}_p(\bh)]=(\phi^{\epsilon}_{ik}\star\phi^{\epsilon}_{pk})(\bh).
\end{align*}
In particular, the field (\ref{eq:AnnMultifractalFieldIndicesIndpt}) converges in a $L_2$ sense when $\epsilon\to 0$, and it has same variance and covariance as the underlying Gaussian field. In the sequel, we note $\bu^{ind}=\lim_{\epsilon\to 0} \bu^{\epsilon,ind}$. 

\subsection{Structure functions}

In a similar way, we define the velocity increment $\delta_\ell u_i$ as
 \begin{equation}\label{eq:AnnVIFieldIndicesIndpt}
\delta_\ell u_i^{ind,\epsilon} = u_i^{ind,\epsilon}(\bell/2) - u_i^{ind,\epsilon}(-\bell/2) = \frac{1}{c_\epsilon}\int\Phi^{\epsilon,\bell}_{ik}(\bz)\left(\e^{\gamma \tX^\epsilon(\bz)}\right)_{kl}W_l(\diff z),
\end{equation}
where we have defined the even function
$$ \Phi^{\epsilon,\bell}_{ik}(\bx)=\phi^\epsilon_{ik}(\bx+\bell/2)-\phi^\epsilon_{ik}(\bx-\bell/2). $$ 

As we have seen, the covariance of the vector field (\ref{eq:AnnMultifractalFieldIndicesIndpt}) is the same as the one obtained from the underlying Gaussian field. Thus it has also the same second order structure functions, both longitudinal and transverse (see annex \ref{ann:GF}). Take $n\in \mathbb N^*$. Using the independence between $\tX^\epsilon$ and $\bW$, we can write (hereafter, no summation over the index $i$),
$$\E (\delta_\ell u_i^{ind,\epsilon})^n = \frac{1}{c_\epsilon^n}\int\prod_{q=1}^n\Phi^{\epsilon,\bell}_{ik_q}(\bs{z_q})\E \left[\prod_{q=1}^n\left(\e^{\gamma \tX^\epsilon(\bs{z_q})}\right)_{k_ql_q}\right]\E \left[\prod_{q=1}^nW_{l_q}(\diff z_q)\right].$$
We can see that all odd-order structure functions vanish since the expectation of an odd product of the vector white noise components is always zero, i.e.
$$ \E \left[\prod_{q=1}^{2n+1}W_{l_q}(\diff z_q)\right]=0.$$
Thus
$$\E (\delta_\ell u_i^{ind,\epsilon})^{2n+1}=0.$$
This field (\ref{eq:AnnMultifractalFieldIndicesIndpt}) is not dissipative, and does not exhibit energy transfer. We consider in the sequel only even-order structure functions to get:
$$\E (\delta_\ell u_i^{ind,\epsilon})^{2n} = \frac{1}{c_\epsilon^{2n}}\int\prod_{q=1}^{2n}\Phi^{\epsilon,\bell}_{ik_q}(\bs{z_q})\E \left[\prod_{q=1}^{2n}\left(\e^{\gamma \tX^\epsilon(\bs{z_q})}\right)_{k_ql_q}\right]\E \left[\prod_{q=1}^{2n}W_{l_q}(\diff z_q)\right].$$
Factorizing the $2n$-product of the white noise vector (Isserlis' theorem) as
\begin{align*}
\E \left[\prod_{q=1}^{2n} W_{l_q}(\diff z_q)\right] & = \frac{1}{2^nn!}\sum_{\mathcal S\in S_{2n}} \prod_{i=1}^n \E \left[W_{l_{\mathcal S(2i-1)}}(\diff z_{\mathcal S(2i-1)})W_{l_{\mathcal S(2i)}}(\diff z_{\mathcal S(2i)})\right]\\
&= \frac{1}{2^nn!}\sum_{\mathcal S\in S_{2n}} \prod_{i=1}^n \delta_{l_{\mathcal S(2i-1)},l_{\mathcal S(2i)}}\delta_{z_{\mathcal S(2i-1)},z_{\mathcal S(2i)}}dz_{\mathcal S(2i)}
\end{align*}
where $S_{2n}$ is the permutation ensemble of $\{1,...,2n\}$, which has cardinality $(2n)!$. Thus
\begin{align*}
&\E (\delta_\ell u_i^{ind,\epsilon})^{2n} = \frac{1}{2^nn!c_\epsilon^{2n}}\sum_{\mathcal S\in
S_{2n}} \\
&\int\prod_{q=1}^{n}\Phi^{\epsilon,\bell}_{ik_{\mathcal S(2q-1)}}(\bs{z_{\mathcal S(2q)}})\Phi^{\epsilon,\bell}_{ik_{\mathcal S(2q)}}(\bs{z_{\mathcal S(2q)}})\E \left[\prod_{q=1}^{n}\left(\e^{2\gamma \tX^\epsilon(\bs{z_{\mathcal S(2q)}})}\right)_{k_{\mathcal S(2q-1)}k_{\mathcal S(2q)}}\right]\prod_{q=1}^n \diff z_{\mathcal S(2q)}.
\end{align*}
It is clear that the sum over the permutations gives $(2n)!$ equal contributions, and we are left with
\begin{align}\label{eq:2nIndptSF}
\E (\delta_\ell u_i^{ind,\epsilon})^{2n} = \frac{(2n)!}{2^nn!c_\epsilon^{2n}}\int\prod_{q=1}^{n}\Phi^{\epsilon,\bell}_{ik_{2q-1}}(\bs{z_{2q}})\Phi^{\epsilon,\bell}_{ik_{2q}}(\bs{z_{2q}})\E \left[\prod_{q=1}^{n}\left(\e^{2\gamma \tX^\epsilon(\bs{z_{2q}})}\right)_{k_{2q-1}k_{2q}}\right]\prod_{q=1}^n \diff z_{2q}.
\end{align}
Note that, in the Gaussian case, i.e. $\gamma=0$, which is equivalent to $\left(\e^{2\gamma \tX^\epsilon(\bs{z_{2q}})}\right)_{k_{2q-1}k_{2q}}=\delta_{k_{2q-1}k_{2q}}$ and $c_\epsilon=1$, we recover the  statistics obtained for the underlying Gaussian field (see annex \ref{ann:GF}), namely
\begin{align*}
\E (\delta_\ell u_i^{ind,\epsilon})^{2n} &\build{=}_{}^{\gamma=0}  \E (\delta_\ell u_i^{\epsilon,g})^{2n}\\
&= \frac{(2n)!}{2^nn!}\int\prod_{q=1}^{n}\Phi^{\epsilon,\bell}_{ik_{2q}}(\bs{z_{2q}})\Phi^{\epsilon,\bell}_{ik_{2q}}(\bs{z_{2q}})\prod_{q=1}^n\diff z_{2q}\\
&= \frac{(2n)!}{2^nn!}\left[\int\Phi^{\epsilon,\bell}_{ik}(\bz)\Phi^{\epsilon,\bell}_{ik}(\bz)\diff z\right]^n\\
&= \frac{(2n)!}{2^nn!}\left[\left(\Phi^{\epsilon,\bell}_{ik}\star\Phi^{\epsilon,\bell}_{ik}\right)(0)\right]^n= \frac{(2n)!}{2^nn!}\left[\E (\delta_\ell u_i^{ind,\epsilon})^{2}\right]^n= \frac{(2n)!}{2^nn!}\left[\E (\delta_\ell u_i^{\epsilon,g})^{2}\right]^n.
\end{align*}

\subsubsection{Fourth order structure functions and flatnesses}

Of special interest are the fourth order structure functions $\E (\delta_\ell u_i^\epsilon)^{4}$ (longitudinal and transverse cases), and the respective flatnesses, i.e.  $\mathcal F^{\parallel}$ and $\mathcal F^{\perp}$ (\ref{eq:Flatnesses}). We get (no summation over repeated index $i$ implied),
\begin{align*}
\E (\delta_\ell & u_i^{\epsilon,ind})^{4} = \\
&\frac{3}{c_\epsilon^{4}}\int\Phi^{\epsilon,\bell}_{ik_{1}}(\bs{z_{2}})\Phi^{\epsilon,\bell}_{ik_{2}}(\bs{z_2})\Phi^{\epsilon,\bell}_{ik_{3}}(\bs{z_4})\Phi^{\epsilon,\bell}_{ik_{4}}(\bs{z_4})\E \left[\left(\e^{2\gamma \tX^\epsilon(\bs{z_2})}\right)_{k_{1}k_{2}}\left(\e^{2\gamma \tX^\epsilon(\bs{z_4})}\right)_{k_{3}k_{4}}\right]\diff z_{2}\diff z_{4}.
\end{align*}
Let us massage slightly the covariance of the matrix exponentials that enters the former expression. To do so, we will use the theory developed by \cite{CheRho13}. The field of isotropic symmetric  matrices $\tX^\epsilon$ is homogeneous, and furthermore, the joint density of two matrices at the locations $\bs{z_2}$ and $\bs{z_4}$ depends only on the distance $|\bs{z_4}-\bs{z_2}|$. Thus, it can be shown that 
$$\frac{1}{c_\epsilon^4}\E \left[\left(\e^{2\gamma \tX^\epsilon(0)}\right)_{k_{1}k_{2}}\left(\e^{2\gamma \tX^\epsilon(\bh)}\right)_{k_{3}k_{4}}\right] = f_\epsilon (h)\delta_{k_1k_2}\delta_{k_3k_4} + g_\epsilon(h)\left[\delta_{k_1k_3}\delta_{k_2k_4}+\delta_{k_1k_4}\delta_{k_2k_3} \right], $$
where
$$ f_\epsilon(h) = \frac{1}{15c_\epsilon^4}\left[ 2\E \left( \tr~\e^{2\gamma \tX^\epsilon(0)}\tr~\e^{2\gamma \tX^\epsilon(\bh)}\right)-\E \left( \tr~\e^{2\gamma \tX^\epsilon(0)}\e^{2\gamma \tX^\epsilon(\bh)}\right)\right] $$
and
$$ g_\epsilon(h) = \frac{1}{30c_\epsilon^4}\left[ 3\E \left( \tr~\e^{2\gamma \tX^\epsilon(0)}\e^{2\gamma \tX^\epsilon(\bh)}\right)-\E \left( \tr~\e^{2\gamma \tX^\epsilon(0)}\tr~\e^{2\gamma \tX^\epsilon(\bh)}\right)\right]. $$
We remark that if $\tX^\epsilon$ is proportional to the identity, the function $g_\epsilon$ vanishes. Using this isotropic form of a fourth-order tensor, we obtain
\begin{align*}
\E (\delta_\ell u_i^{ind,\epsilon})^{4} &= 3\int\Phi^{\epsilon,\bell}_{ip}(\bs{z_2})\Phi^{\epsilon,\bell}_{ip}(\bs{z_2})\Phi^{\epsilon,\bell}_{iq}(\bs{z_4})\Phi^{\epsilon,\bell}_{iq}(\bs{z_4})f_\epsilon(|\bs{z_2}-\bs{z_4}|)\diff z_{2}\diff z_{4}\\
 &+ 6\int\Phi^{\epsilon,\bell}_{ip}(\bs{z_2})\Phi^{\epsilon,\bell}_{iq}(\bs{z_2})\Phi^{\epsilon,\bell}_{ip}(\bs{z_4})\Phi^{\epsilon,\bell}_{iq}(\bs{z_4})g_\epsilon(|\bs{z_2}-\bs{z_4}|)\diff z_{2}\diff z_{4}.
\end{align*}
Using the results of \cite{CheRho13}, we get
\begin{align*}
\E \left[ \tr~\e^{2\gamma \tX^\epsilon(0)}\e^{2\gamma \tX^\epsilon(\bh)}\right]\build{\sim}_{\epsilon\to 0}^{}4(2\gamma\sigma_\epsilon)^4\left(1+\frac{1}{2}\right)^2\e^{4\gamma^2\sigma_\epsilon^2}\e^{-4\frac{1}{2}\gamma^2\sigma_{h}^2}\int_{O_3(\mathbb R)}|O_{11}|^2\e^{4(1+\frac{1}{2})\gamma^2\sigma^2_{h}|O_{11}|^2}\diff O,
\end{align*}
and similarly,
\begin{align*}
\E \left[ \tr~\e^{2\gamma \tX^\epsilon(0)}\tr~\e^{2\gamma \tX^\epsilon(\bh)}\right]\build{\sim}_{\epsilon\to 0}^{}4(2\gamma\sigma_\epsilon)^4\left(1+\frac{1}{2}\right)^2\e^{4\gamma^2\sigma_\epsilon^2}\e^{-4\frac{1}{2}\gamma^2\sigma_{h}^2}\int_{O_3(\mathbb R)}\e^{4(1+\frac{1}{2})\gamma^2\sigma^2_{h}|O_{11}|^2}\diff O,
\end{align*}
where the remaining integration is performed over the orthogonal group $O_3(\mathbb R)$, and recall that asymptotically $\sigma^2_\epsilon \sim \ln L/\epsilon$ (when $\epsilon\to 0$) and $\sigma^2_{h} \sim \ln L/h$ (when $\epsilon\to 0$ and after $h\to 0$). In the same fashion, we can find
$$c_\epsilon^2=\E \left[ \frac{1}{3}\tr~\e^{2\gamma \tX^\epsilon}\right]\build{\sim}_{\epsilon\to 0}^{} \frac{8}{3}\gamma^2\sigma_\epsilon^2\left(1+\frac{1}{2}\right)\e^{2\gamma^2\sigma_\epsilon^2}.$$
Thus,
\begin{align*}
\lim_{\epsilon\to 0} \frac{1}{c_\epsilon^4}\E \left[ \tr~\e^{2\gamma \tX^\epsilon(0)}e^{2\gamma \tX^\epsilon(\bh)}\right]=3^2\e^{-4\frac{1}{2}\gamma^2\sigma_{h}^2}\int_{O_3(\mathbb R)}|O_{11}|^2\e^{4(1+\frac{1}{2})\gamma^2\sigma^2_{h}|O_{11}|^2}\diff O,
\end{align*}
and similarly,
\begin{align*}
\lim_{\epsilon\to 0}\frac{1}{c_\epsilon^4}\E \left[ \tr~\e^{2\gamma \tX^\epsilon(0)}\tr~\e^{2\gamma \tX^\epsilon(\bh)}\right] =3^2\e^{-4\frac{1}{2}\gamma^2\sigma_{h}^2}\int_{O_3(\mathbb R)}e^{4(1+\frac{1}{2})\gamma^2\sigma^2_{h}|O_{11}|^2}\diff O.
\end{align*}
This shows that the quantities $f_\epsilon$ and $g_\epsilon$ converge when $\epsilon\to 0$. We will note $f$ and $g$ their respective limits. Thus,
\begin{align*}
f(h)=\frac{3^2}{15}\e^{-4\frac{1}{2}\gamma^2\sigma_{h}^2}\left[2\int_{O_3(\mathbb R)}e^{4(1+\frac{1}{2})\gamma^2\sigma^2_{h}|O_{11}|^2}\diff O-\int_{O_3(\mathbb R)}|O_{11}|^2\e^{4(1+\frac{1}{2})\gamma^2\sigma^2_{h}|O_{11}|^2}\diff O\right],
\end{align*}
and 
\begin{align*}
g(h)=\frac{3^2}{30}\e^{-4\frac{1}{2}\gamma^2\sigma_{h}^2}\left[3\int_{O_3(\mathbb R)}|O_{11}|^2e^{4(1+\frac{1}{2})\gamma^2\sigma^2_{h}|O_{11}|^2}\diff O-\int_{O_3(\mathbb R)}e^{4(1+\frac{1}{2})\gamma^2\sigma^2_{h}|O_{11}|^2}\diff O\right].
\end{align*}
Introducing the Dawson integral $\mathcal G$
$$ z\in \mathbb R \mapsto \mathcal G (z) = \e^{-z^2} \int_{0}^z e^{y^2}dy$$
we can obtain an explicit expression of former angular integrals at a finite scale $\ell$ as
\begin{align}\label{eq:fFiniteSize}
f(h)=\frac{1}{5}\frac{\e^{4\gamma^2\sigma_{h}^2}}{4\gamma^2\sigma^2_{h}}\left[ 4\sqrt{6}\gamma \sigma_{h}\mathcal G(\sqrt{6}\gamma \sigma_{h})-\frac{\sqrt{6}\gamma \sigma_{h}-\mathcal G(\sqrt{6}\gamma \sigma_{h})}{\sqrt{6}\gamma \sigma_{h}} \right],
\end{align}
and 
\begin{align}\label{eq:gFiniteSize}
g(h)=\frac{1}{5}\frac{\e^{4\gamma^2\sigma_{h}^2}}{4\gamma^2\sigma^2_{h}}\left[\frac{3}{2}\frac{\sqrt{6}\gamma \sigma_{h}-\mathcal G(\sqrt{6}\gamma \sigma_{h})}{\sqrt{6}\gamma \sigma_{h}} -\sqrt{6}\gamma \sigma_{h}\mathcal G(\sqrt{6}\gamma \sigma_{h}) \right].
\end{align}
Using the asymptotic behaviour of the Dawson integral, namely $2x\mathcal G(x) \to 1$ when $x\to\infty$, we get the asymptotic behaviour of the functions $f$ and $g$, that is
$$f(h)\build{\sim}_{h\to 0}^{}g(h)\build{\sim}_{h\to 0}^{}\frac{1}{5}\frac{\e^{4\gamma^2\sigma_{h}^2}}{4\gamma^2\sigma^2_{h}}.$$
Recall that asymptotically, the covariance of the diagonal elements of $\tX$ is logarithmic (see equation \ref{eq:AsympSigmaCorr}). We can thus write it as
\begin{equation}\label{eq:AsympSigmaCorrAlpha}
\sigma^2_{h} = \ln\left(\frac{L}{h}\right) + \alpha(h),
\end{equation}
where $\alpha(h)$ is a bounded function of its argument, showing that
$$f(h)\build{\sim}_{h\to 0}^{}g(h)\build{\sim}_{h\to 0}^{}\frac{1}{5}\frac{\e^{4\gamma^2\alpha(0)}}{4\gamma^2 \ln\left(\frac{L}{h}\right)}\left( \frac{L}{h}\right)^{4\gamma^2}.$$
If the integrals exist, we can obtain the limit $\epsilon\to 0$ of the fourth order moment of velocity increments as
$$\E (\delta_\ell u^{ind}_i)^{4} = \lim_{\epsilon \to 0}\E (\delta_\ell u_i^{ind,\epsilon})^{4}, $$
with
\begin{align*}
\E (\delta_\ell u^{ind}_i)^{4} &= 3\int\Phi^{\bell}_{ip}(\bs{z_2})\Phi^{\bell}_{ip}(\bs{z_2})\Phi^{\bell}_{iq}(\bs{z_4})\Phi^{\bell}_{iq}(\bs{z_4})f(|\bs{z_2}-\bs{z_4}|)\diff z_{2}\diff z_{4}\\
 &+ 6\int\Phi^{\bell}_{ip}(\bs{z_2})\Phi^{\bell}_{iq}(\bs{z_2})\Phi^{\bell}_{ip}(\bs{z_4})\Phi^{\bell}_{iq}(\bs{z_4})g(|\bs{z_2}-\bs{z_4}|)\diff z_{2}\diff z_{4}\\
&= 3\int (\Phi^{\bell}_{ip}\Phi^{\bell}_{ip}\star \Phi^{\bell}_{iq}\Phi^{\bell}_{iq})(\bh)f(h)\diff h+6\int (\Phi^{\bell}_{ip}\Phi^{\bell}_{iq}\star \Phi^{\bell}_{ip}\Phi^{\bell}_{iq})(\bh)g(h)\diff h.
\end{align*}
We must still make sure that the integrals exist. Clearly, the strongest singularities are obtained when the integration variable $h$ is such that its modulus is zero. While performing the integration over $\bh$ in the asymptotic form of $\E (\delta_\ell u_i)^{4}$, we thus encounter the singularity $1/h^{4\gamma^2+3-2H}$ which is integrable in three dimensions only if 
$$ 2\gamma^2<H.$$
Let us get the scaling behaviour of the fourth order structure function. Define first the unit-vector $\bs{e}$ and write $\bell = \ell \bs{e}$. Recall too that (see the scaling arguments developed in annex \ref{ann:GF})
$$\Phi^{\bell}_{ip}(\ell \bz) \build{\sim}_{\ell\to 0}^{}\ell^{H-\frac{3}{2}}\tilde{\Phi}^{\bs{e}}_{ip}(\bz) $$
with 
$$\tilde{\Phi}^{\bs{e}}_{ip}(\bz) =\epsilon_{ijp}\varphi_L(0)\left[\frac{z_j+e_j/2}{|\bz+\bs{e}/2|^{\frac{5}{2}-H}}+\frac{e_j/2-z_j}{|\bs{e}/2-\bx|^{\frac{5}{2}-H}}\right].$$
In the same spirit, the intermittent correction to the fourth order structure function comes from the asymptotic behaviour of $f$ and $g$ at small arguments, namely
$$f(\ell h)\build{\sim}_{\ell\to 0}^{}g(\ell h)\build{\sim}_{\ell\to 0}^{}\frac{1}{5}\frac{1}{4\gamma^2\ln\frac{1}{\ell}}\left( \frac{L}{h}\right)^{4\gamma^2}\left( \frac{1}{\ell}\right)^{4\gamma^2}\e^{4\gamma^2\alpha(0)} .$$
We thus obtain the following scaling behaviour of the fourth order structure function
\begin{align*}
\E (\delta_\ell u^{ind}_i)^{4} \build{\sim}_{\ell\to 0}^{} \frac{\ell^{4H-4\gamma^2}}{20\gamma^2\ln\frac{1}{\ell}}e^{4\gamma^2\alpha(0)}\int \left[3(\tilde{\Phi}^{\bs{e}}_{ip}\tilde{\Phi}^{\bs{e}}_{ip}\star \tilde{\Phi}^{\bs{e}}_{iq}\tilde{\Phi}^{\bs{e}}_{iq})+6 (\tilde{\Phi}^{\bs{e}}_{ip}\tilde{\Phi}^{\bs{e}}_{iq}\star \tilde{\Phi}^{\bs{e}}_{ip}\tilde{\Phi}^{\bs{e}}_{iq})\right](\bh)\left( \frac{L}{h}\right)^{4\gamma^2}\diff h.
\end{align*}
This shows that longitudinal and transverse fourth order structure functions have the same scaling behaviours. More precisely, without loss of generality, consider the first velocity component $u_1$ and the two unit vectors $\bs{e}=\bs{e_1}=(1,0,0)$ and $\bs{e}=\bs{e_2}=(0,1,0)$ for the transverse one, we finally get
$$ \mathbb E (\delta_\ell^{\parallel} u^{ind})^4 \build{\sim}_{\ell\rightarrow 0}^{} C_4^{ind,\parallel}\frac{\ell^{4H-4\gamma^2}}{\ln\frac{1}{\ell}} \, \mbox{ and } \, \mathbb E (\delta_\ell^{\perp} u^{ind})^4 \build{\sim}_{\ell\rightarrow 0}^{} C_4^{ind,\perp}\frac{\ell^{ 4H-4\gamma^2}}{\ln\frac{1}{\ell}}$$
with 
$$C_4^{ind,\parallel} = \frac{\e^{4\gamma^2\alpha(0)}}{20\gamma^2}\int \left[3(\tilde{\Phi}^{\bs{e_1}}_{1p}\tilde{\Phi}^{\bs{e_1}}_{1p}\star \tilde{\Phi}^{\bs{e_1}}_{1q}\tilde{\Phi}^{\bs{e_1}}_{1q})+6 (\tilde{\Phi}^{\bs{e_1}}_{1p}\tilde{\Phi}^{\bs{e_1}}_{1q}\star \tilde{\Phi}^{\bs{e_1}}_{1p}\tilde{\Phi}^{\bs{e_1}}_{1q})\right](\bh)\left( \frac{L}{h}\right)^{4\gamma^2}\diff h$$ 
and
$$C_4^{ind,\perp} = \frac{e^{4\gamma^2\alpha(0)}}{20\gamma^2}\int \left[3(\tilde{\Phi}^{\bs{e_2}}_{1p}\tilde{\Phi}^{\bs{e_2}}_{1p}\star \tilde{\Phi}^{\bs{e_2}}_{1q}\tilde{\Phi}^{\bs{e_2}}_{1q})+6 (\tilde{\Phi}^{\bs{e_2}}_{1p}\tilde{\Phi}^{\bs{e_2}}_{1q}\star \tilde{\Phi}^{\bs{e_2}}_{1p}\tilde{\Phi}^{\bs{e_2}}_{1q})\right](\bh)\left( \frac{L}{h}\right)^{4\gamma^2}\diff h \mbox{ .}$$ 
This shows that the velocity field $\bu^{ind}$ (\ref{eq:MultifractalFieldIndicesIndpt}), built assuming independence of $\tX$ and $\bW$, is intermittent, and the respective flatnesses (\ref{eq:Flatnesses}) behave as power-laws times a logarithmic correction with (see annex \ref{ann:GF} for the expression of the second-order structure functions that are the same as the ones obtained from the underlying Gaussian velocity field)
\begin{equation}\label{eq:AnnPredIndFlat}
\mathcal F^{\parallel}(\ell)\build{\sim}_{\ell\to 0}^{} \frac{C_4^{ind,\parallel}}{(C_2^{ind,\parallel})^2} \frac{\ell^{-4\gamma^2}}{\ln\frac{1}{\ell}} \, \mbox{ and } \, \mathcal F^{\perp}(\ell)\build{\sim}_{\ell\to 0}^{} \frac{C_4^{ind,\perp}}{(C_2^{ind,\perp})^2} \frac{\ell^{-4\gamma^2}}{\ln\frac{1}{\ell}}. 
\end{equation}




\end{document}